\documentclass[]{aa}
\input psfig.sty

\newcommand{\lsun}{$log L/L_{\odot}\,$}
\newcommand{\msun}{$M/M_{\odot}\,$}
\newcommand{\pve}{$P_2 / P_0 \,$}
\newcommand{\cw}{{\em Christy wave}}

\begin{document}
   \thesaurus{06         
              (08.04.1;  
               08.05.3;  
               08.15.1;  
               08.22.1;
               11.13.1;
               02.08.1)} 
\title{Classical Cepheid pulsation models. VI. The Hertzsprung progression} 

   \author{G. Bono 
           \inst{1}
   \and    M. Marconi  
          \inst{2}
   \and    R. F. Stellingwerf  
	  \inst{3}	 
          }
          
\institute{Osservatorio Astronomico di Roma, Via Frascati 33, 00040 Monte 
Porzio Catone, Italy;  bono@coma.mporzio.astro.it  
\and Osservatorio Astronomico di Capodimonte, Via Moiariello 16, 80131 Napoli, 
Italy; marcella@na.astro.it 
\and SC, 2229 Loma Linda, Los Alamos, NM 87544, USA; rfs@stellingwerf.com 
}

\offprints{G. Bono: bono@coma.mporzio.astro.it}


\titlerunning{The Hertzsprung progression} 

\authorrunning{Bono et al. } 

\maketitle

\begin{abstract}

We present the results of an extensive theoretical investigation on 
the pulsation behavior of Bump Cepheids. We constructed several sequences 
of full amplitude, nonlinear, convective models by adopting a chemical 
composition typical of Large Magellanic Cloud (LMC) Cepheids 
(Y=0.25, Z=0.008) and stellar masses ranging from \msun=6.55 to 7.45. 
We find that theoretical light and velocity curves 
reproduce the HP, and indeed close to the blue edge the bump is located 
along the descending branch, toward longer periods it crosses at first 
the luminosity/velocity maximum and then it appears along the rising 
branch. In particular, we find that the predicted period at the HP 
center is $P_{\rm {HP}}=11.24\pm0.46$ d and that such a value is in very 
good agreement with the empirical value estimated by adopting the 
Fourier parameters of LMC Cepheid light curves i.e. 
$P_{\rm{HP}}=11.2\pm0.8$ d (Welch et al. 1997). 
Moreover, light and velocity amplitudes present a "double-peaked" 
distribution which is in good qualitative agreement with observational 
evidence on Bump Cepheids. 
It turns out that both the skewness and the acuteness typically 
show a well-defined minimum at the HP center and the periods 
range from $P_{\rm{HP}}=10.73\pm0.97$ d to $P_{\rm{HP}}=11.29\pm0.53$ d 
which are in good agreement with empirical estimates. We also 
find that the models at the HP center are located 
within the resonance region but not on the 2:1 resonance line 
(\pve $=0.5$), and indeed the \pve ratios roughly range 
from 0.51 (cool models) to 0.52 (hot models).  

Interestingly enough, the predicted Bump Cepheid masses, based on a 
Mass-Luminosity (ML) relation which neglects the convective core 
overshooting, are in good agreement with the empirical masses of 
Galactic Cepheids estimated by adopting the Baade-Wesselink method 
(Gieren 1989). As a matter of fact, the observed mass at the HP 
center -$P\approx11.2$ d- is $6.9\pm0.9\; M_\odot$, while the 
predicted mass is $7.0\pm0.45\; M_\odot$. 
Even by accounting for the metallicity difference between Galactic 
and LMC Cepheids, this result seems to settle down the long-standing 
problem of the Bump mass discrepancy.  
 
Finally, the dynamical behavior of a cool Bump Cepheid model provides a plain 
explanation of an ill-understood empirical evidence. In fact, it turns out 
that toward cooler effective temperatures the bump becomes the main maximum, 
while the true maximum is the bump which appears along the rising branch.  
This finding also supplies a plain explanation of the reason why the 
pulsation amplitudes of Bump Cepheids present a "double-peaked" 
distribution. 


\end{abstract}

\section{INTRODUCTION}

More than seventy years ago Hertzsprung (1926) discovered that a 
subsample of Galactic classical Cepheids presents a relationship 
between the bump along the light curve and the pulsation period; 
the so-called "Hertzsprung Progression" was subsequently 
discovered among Magellanic Clouds (MCs) and Andromeda Cepheids by 
Kukarkin \& Parenago (1937), Shapley \& Mckibben Nail (1940), and 
by Payne-Gaposchkin (1951,1954). The HP observational scenario was 
enriched by Joy (1937) and by Ledoux \& Walraven (1958) who found 
a similar behavior in radial velocity curves.  
 
The empirical finger-print of the HP is the following:
classical Cepheids in the period range $6 < P < 16$ d 
show a bump along both the light and the velocity curves. 
This secondary feature appears on the descending branch of the 
light curve for Cepheids with periods up to 9 days, while it appears 
close to maximum light for $9 < P < 12 $ d and moves at earlier 
phases for longer periods. On the basis of this observational evidence  
this group of variables was christened "Bump Cepheids" for avoiding 
to be mixed-up with "Beat Cepheids". In fact, the latter group refer 
to mixed-mode variables -i.e. objects in which two or more modes are 
simultaneously excited- and therefore both the shape of the light 
curves and the pulsation amplitudes change from one cycle to the 
next, whereas Bump Cepheids are single mode variables and their 
pulsation properties are characterized by a strong regularity over 
consecutive cycles.  
 
A more quantitative approach concerning Bump Cepheids was originally 
suggested by Kukarkin \& Parenago (1937), Payne-Gaposchkin (1947), 
and more recently by Simon \& Lee (1981) who investigated the shape of 
the light curves by means of Fourier analysis. The last authors found that 
both the phase difference -$\phi_{21}$- and the amplitude ratio -$R_{21}$- 
show a sharp minimum close to the HP center. 
Following this approach several investigations have been already devoted 
to Fourier parameters of Galactic and Magellanic Cepheids. 
In particular, Moskalik et al. (1992, hereinafter MBM) 
suggested that the minimum in the Fourier parameters for Galactic 
Cepheids takes place at $P_{\rm{HP}}=10.0 \pm 0.5$ d, while 
Moskalik et al. (2000) by investigating a sample of more than 100 
radial velocity curves found $P_{\rm{HP}}=9.95 \pm 0.05$ d. 
At the same time, Welch et al. (1997) by investigating 
a large sample of Cepheids in the LMC estimated that the 
minimum in the Fourier parameters is located at $P_{\rm{HP}}=11.2 \pm 0.8$ d.  
Thus supporting the shift of the HP center toward longer periods 
originally suggested by Payne-Gaposchkin (1951) and strengthened 
by Andreasen \& Petersen (1987) and by Andreasen (1988). 
More recently Beaulieu (1998) suggested that the HP center in 
LMC and in Small Magellanic Cloud (SMC) Cepheids is located 
at $P_{\rm{HP}}=10.5 \pm 0.5$ d and $P_{\rm{HP}}=11.0 \pm 0.5$ d respectively. 
Since these three stellar systems are characterized by different 
mean metallicities, namely Z=0.02 (Galaxy), Z=0.008 (LMC), and 
Z=0.004 (SMC), this empirical evidence seems to suggest  that a 
decrease in metallicity moves the HP center toward longer periods.  

Up to now, two distinct models have been proposed in the literature
to explain the appearance of the HP among Bump Cepheids; the
{\em echo model} and the {\em resonance model}. The former was
suggested by Whitney (1956) and discussed by Christy (1968,1975)
on the basis of Cepheid nonlinear, radiative models. According to 
Christy, during each cycle  close to the phases of minimum radius 
and before the phase of maximum expansion velocity a pressure excess
is generated in the first He ionization region. This pressure
excess causes a rapid expansion which in turn generates two pressure
waves moving outward and inward. The latter reaches the stellar core
close to the phase of maximum radius, then reflects and reaches the
surface one cycle later causing the appearance of the bump.
The {\em resonance model} was suggested by Simon \& Schmidt (1976,
hereinafter SS) and is based on linear, adiabatic periods. In this
theoretical framework the bump would be caused by a resonance between
the second overtone and the fundamental mode and it takes place when the
period ratio between these two modes is close to 0.5. In particular,
they suggested that the instability of the fundamental mode drives,
due to a resonance, the second overtone instability. This explanation
lies on the evidence that the nonlinear, radiative models constructed
by Stobie (1969) show a bump along the radial velocity curves close to
the resonance line \pve$=0.5$.

Such an extensive observational and theoretical effort devoted 
to Bump Cepheids were not only aimed at understanding the HP but also 
at providing independent estimates of both the mass and the radius 
of these variables. In fact, dating back to Christy (1968, 1975), 
Stobie (1969) and Fricke et al. (1972, hereinafter FSS) 
it was suggested that these two evolutionary parameters can be 
constrained on the basis of period and phase of the bump. 
A different method to estimate the mass, 
based on period ratios, was suggested by Petersen (1973). 
Note that mass determinations based on these two methods present a 
compelling feature: they are based on observables such as periods and 
phases of the bump which are not affected by systematic empirical 
uncertainties and therefore they are only limited by photometric accuracy.  
However, pulsational masses based on these methods are,
with few exceptions (Carson \& Stothers 1988, hereinafter CS), 
systematically smaller than the evolutionary masses. This longstanding 
puzzle raised the so-called Bump mass discrepancy (see also Cox 1980) 
and at the same time supported the use of a ML 
relation based on evolutionary models which include a mild or a 
strong convective core overshooting (Simon 1995; Wood 1998).  
Even though, the new radiative opacities settled down this
{\em vexata questio} (MBM; Kanbur \& Simon 1994; Simon \&
Kanbur 1994, hereinafter SK), recent linear (Buchler et al. 1996;
Simon \& Young 1997) and nonlinear (Wood et al. 1997)
predictions for Cepheids in the MCs present a small discrepancy
with the ML relations predicted by current evolutionary models. 

The main aim of this paper is to use up-to-date nonlinear hydrodynamical 
models which include the coupling between pulsation and convection as 
well as canonical evolutionary masses, to account for the observed 
properties of Bump Cepheids. In this investigation we focus our 
attention on the classical HP and refer to a forthcoming investigation 
(Bono et al. in preparation) a more detailed discussion on the physical mechanisms 
which trigger the appearance of the HP among Bump Cepheids and on other 
resonances recently proposed for both short (Antonello \& Poretti 1986; 
Buchler et al. 1996) and long-period Cepheids (Pel 1978; Antonello 1998). 
In order to supply a detailed theoretical scenario we constructed a 
fine grid of full amplitude, nonlinear, convective 
models by adopting stellar masses ranging from 6.4 to 7.6 $M_\odot$ 
and a fixed chemical composition, namely Y=0.25, Z=0.008. 
We adopted this chemical composition, because accurate observational 
data on LMC Cepheids are currently available in the literature. 
At the same time, this metal abundance can supply useful constraints
on the intrinsic accuracy of our nonlinear models, and indeed recent 
nonlinear, Cepheid models constructed by adopting similar treatments 
of the coupling between pulsation and convection show at intermediate 
metal contents very large pulsation destabilizations and it has been 
suggested that a powerful dissipation mechanism was not properly 
included in current pulsation codes (Buchler 2000).  

In \S 2 we briefly recall the theoretical framework adopted for 
constructing Cepheid models and present nonlinear observables
predicted by these models as well as their light and velocity curves. 
The systematic behavior of both luminosity and velocity amplitudes 
inside the instability strip is investigated in \S 3.  
In this section we also present the skewness and the acuteness 
of light and velocity curves and discuss the use of these parameters 
to mark the position of the HP center. New analytical relations which 
connect the period at the HP center to the stellar mass and to the 
effective temperature of Bump Cepheid models at minimum amplitude 
are also provided in this section. 
In \S 4 we discuss the location of these models in the HR diagram,
and compare the HP center predicted by nonlinear models with linear
and nonlinear resonance lines.    
The dynamical behavior of two Bump Cepheid models located close 
to the HP center and to the red edge of the instability strip  
is investigated in detail in sect. 5. 
The main results of this investigation are summarized in \S 6 together 
with a brief discussion on the observables which can supply tight 
constraints on the accuracy of theoretical models.

\section{Theoretical framework}

The theoretical framework adopted for constructing full amplitude, 
nonlinear, convective Cepheid models was already described in 
Bono et al. (1998) and in paper I (Bono et al. 1999) and therefore it 
is not discussed further here. The reader interested in the physical 
assumptions adopted to account for the coupling between 
pulsation and convection is referred to Buchler (2000). 
In paper III (Bono et al. in preparation) we found that the 
sequence of models constructed by assuming \msun=7.0 and Y=0.25, Z=0.008 
showed a bump along both light and velocity curves which appears at 
earlier pulsation phases when moving toward longer periods. Accordingly,  
the amplitudes of these models showed a "double-peaked" distribution 
with the two maxima located close to the blue (hot) and to the red 
(cool) edge of the instability strip and a well-defined minimum close 
to $P \approx10.7$ d. On the basis of these features we concluded that 
this sequence of models undergoes the HP when moving from the blue to 
the red edge. A similar behavior in the distribution of radial velocity 
amplitudes was also found by CS in a nonlinear, radiative investigation 
of Bump Cepheid models. However as noted by the same authors, the velocity 
amplitudes were systematically larger than the observed ones and the 
predicted periods at the HP center were 10\% longer than expected. 
 
In order to supply a more detailed mapping of the HP inside the 
instability strip we implemented the sequence at \msun=7.0 with 
new series of models 
constructed by adopting a mass step of 0.15 $M_\odot$. Linear and 
nonlinear calculations were performed by adopting the same input 
physics (opacity, equation of state) and the same ML relation adopted 
in our previous investigations and based on evolutionary models which 
neglect the convective core overshooting (see paper I and III for 
more details). The new sequences were extended in mass until we found 
a well-defined minimum in the pulsation amplitudes inside the instability 
strip. For providing a robust relationship between the bump progression 
and the pulsation period we adopted a temperature step of 100 K 
throughout the instability strip and of only 50 K where the bump 
crosses the luminosity maximum and moves from the descending to the 
rising branch. We end up with six new series of models ranging 
in mass from 6.55 to 7.45 $M_\odot$ which show the typical behavior
of the HP. After the initial perturbation of the static model 
(see paper I) the approach to the nonlinear limit cycle required 
a direct time integration ranging from 1500 to 5000 cycles.  

The input parameters of each sequence are listed in Table 1. 
This table gives the nonlinear pulsation properties of the models 
which present a stable limit cycle. We typically adopted a temperature 
step of 100 K, therefore the edges of the instability region can be 
estimated by increasing/decreasing the effective temperature of the 
hottest/coolest model by 50 K. The first three 
columns list for each model the stellar mass (solar units), the 
logarithmic luminosity (solar units), and the static effective 
temperature (K). Columns 4) and 5) give the nonlinear, fundamental 
period (d) and the logarithmic mean radius (solar units). 
The observables listed in the other columns are the following:   
6) fractional radius oscillation, i.e.
$\Delta R/R_{\rm {ph}}\,=\, (R^{\rm{max}}\,-\,R^{\rm {min}}) / R_{\rm {ph}}$ where $R_{\rm {ph}}$
is the photospheric radius;
7) radial velocity amplitude (km $s^{-1}$), i.e.
$\Delta u \,=\, u^{\rm{max}}\,-\,u^{\rm {min}}$;
8) bolometric amplitude (mag), i.e.
$\Delta M_{\rm {bol}} \,=\, M_{\rm {bol}}^{\rm{max}}\,-\,M_{\rm {bol}}^{\rm {min}}$;
9) logarithmic amplitude of static gravity, i.e.
$\Delta log g_{\rm s} \,=\, log g_{\rm s}^{\rm{max}}\,-\, log g_{\rm s}^{\rm {min}}$;
10) logarithmic amplitude of effective gravity, i.e.
$\Delta log g_{\rm {eff}} \,=\, log g_{\rm {eff}}^{\rm{max}}\,-\,log g_{\rm {eff}}^{\rm {min}}$
where $g_{\rm {eff}}  \,=\, GM/R^2\,+\, du/dt$;
11) temperature amplitude (K), i.e.
$\Delta T \,=\, T^{\rm{max}}\,-\,T^{\rm {min}}$ where $T$ is the temperature
of the outer boundary;
12) effective temperature amplitude (K), i.e.
$\Delta T_{\rm e} \,=\, T_{\rm e}^{\rm{max}}\,-\,T_{\rm e}^{\rm {min}}$ where $T_{\rm e}$ is
derived from the surface luminosity, 
13) total Kinetic energy (erg). 
The quantities listed in columns 4) to 11) refer, with the exception 
of the effective gravity, to the surface zone. 

\begin{figure*}[h]
\psfig{figure=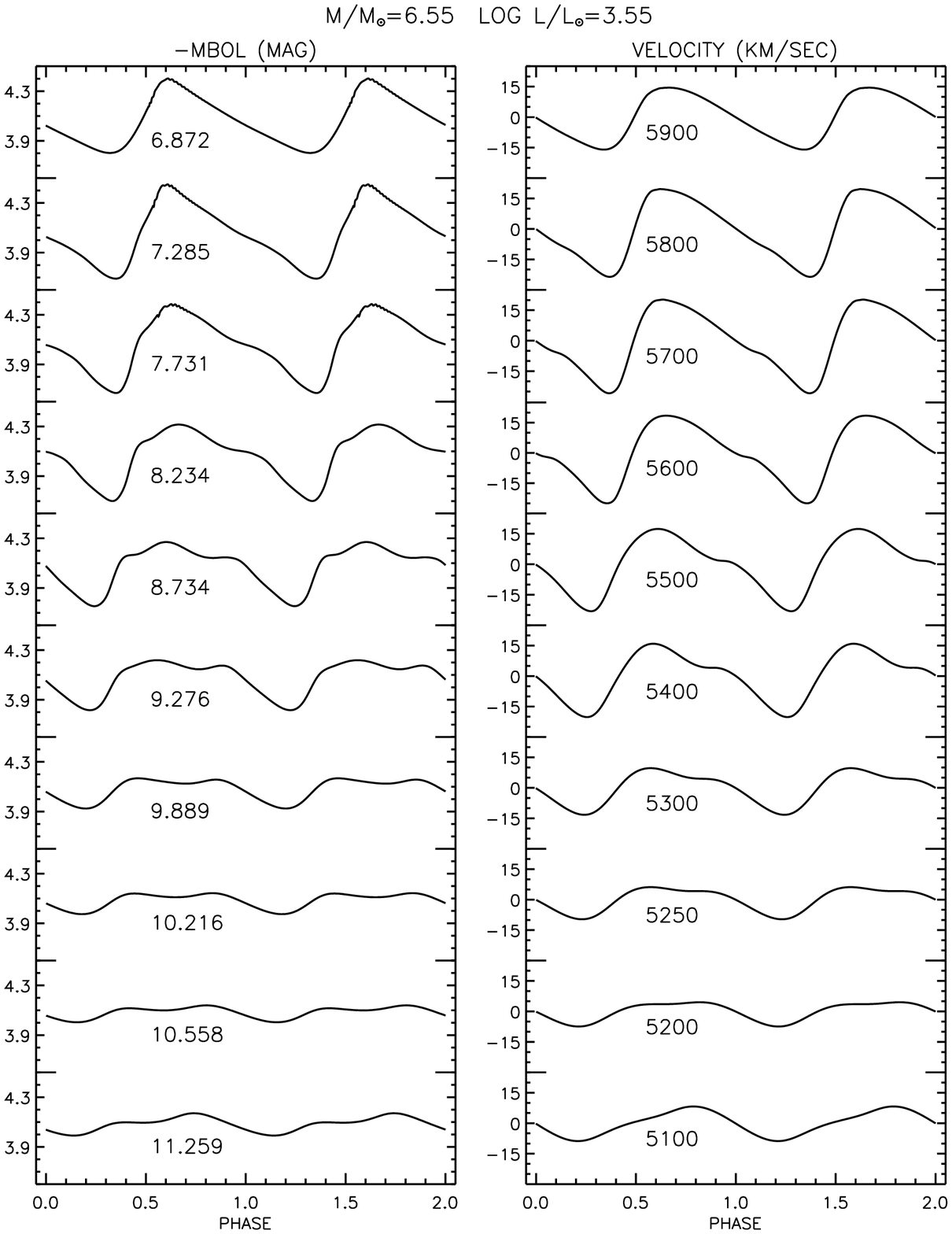,height=15cm,width=18cm} 
\caption {Bolometric light curves (left panel) and radial velocity
(right panel) curves along two consecutive cycles as a function 
of the pulsation phase. The curves plotted in this figure refer 
to a selection of models constructed at fixed mass and luminosity 
(see labeled values). The nonlinear periods (d), and the effective 
temperatures (K) are listed in left and right panels. Positive and 
negative values denote expansion and contraction phases, respectively.}
\end{figure*}

Figs. 1-6, show the light and the velocity curves of the six new sequences 
for the labeled values of both mass and luminosities. Light and velocity 
curves of the sequence for \msun =7.0 were already published in the 
ApJ on-line edition of paper III (Fig. 11f). Note that light and velocity 
curves plotted in these figures show neither spurious secondary features 
(bumps or dips) along the pulsation cycle nor sudden jumps close to the 
phases of maximum compression. As already noted in paper I and III, 
this feature is a significative improvement in comparison with light 
and velocity curves predicted by nonlinear, radiative models 
(Christy 1975), and indeed theoretical curves were plotted without 
applying any running average or filtering process for smoothing 
the surface variations (Karp 1975). This finding is even more 
relevant for investigating the HP. In fact, up to now nonlinear, 
radiative predictions were mainly based on radial velocity curves, since 
the light curves presented several spurious secondary features (CS; MBM). 
As a consequence, even though the bulk of observational data on 
the HP comes from light curves, theoretical insights were focused 
on the shape of radial velocity curves. 

\begin{figure*}[h] 
\psfig{figure=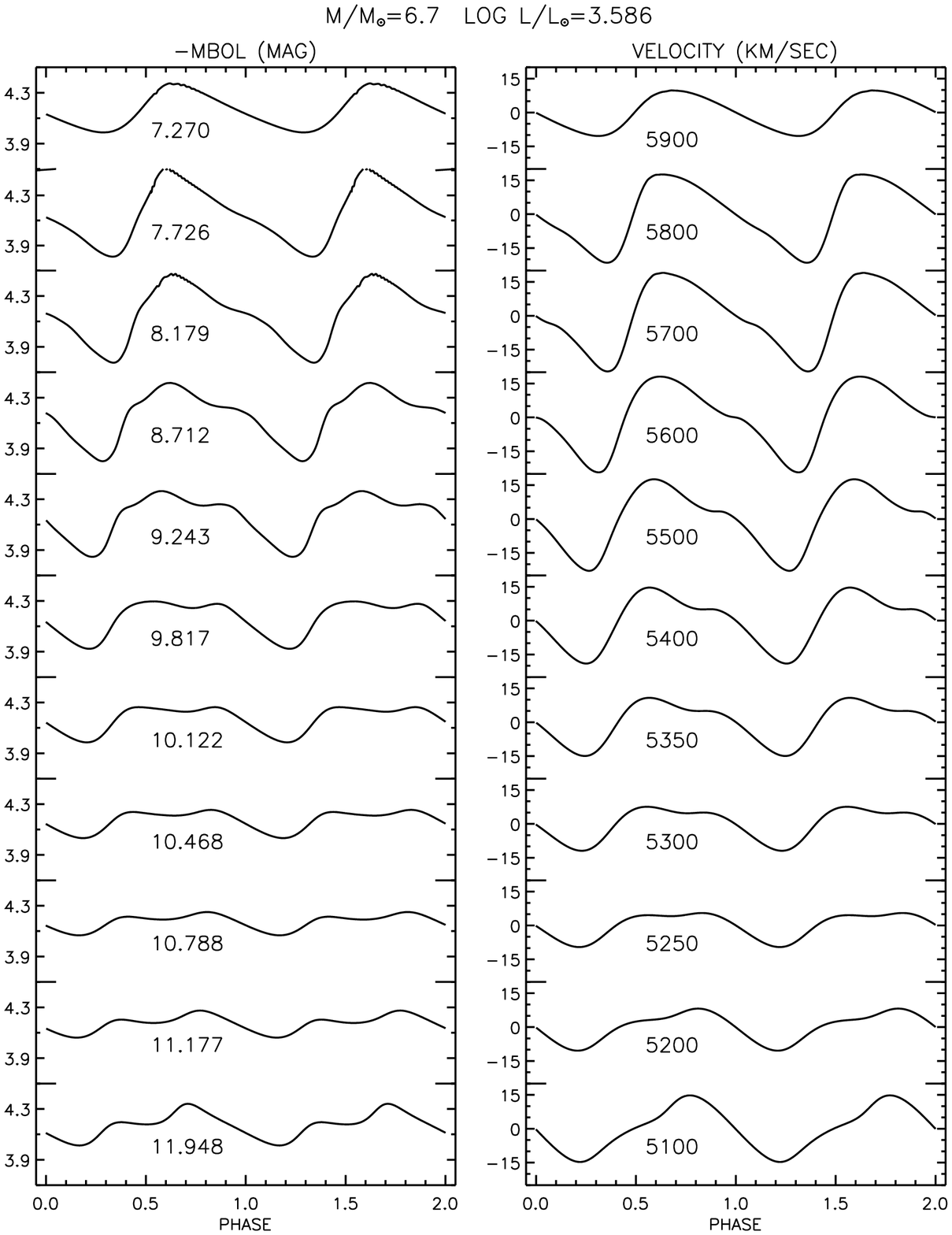,height=15cm,width=18cm} 
\caption {Similar to Fig. 1, but for \msun=6.7 models.}
\end{figure*} 

\begin{figure*}[h] 
\psfig{figure=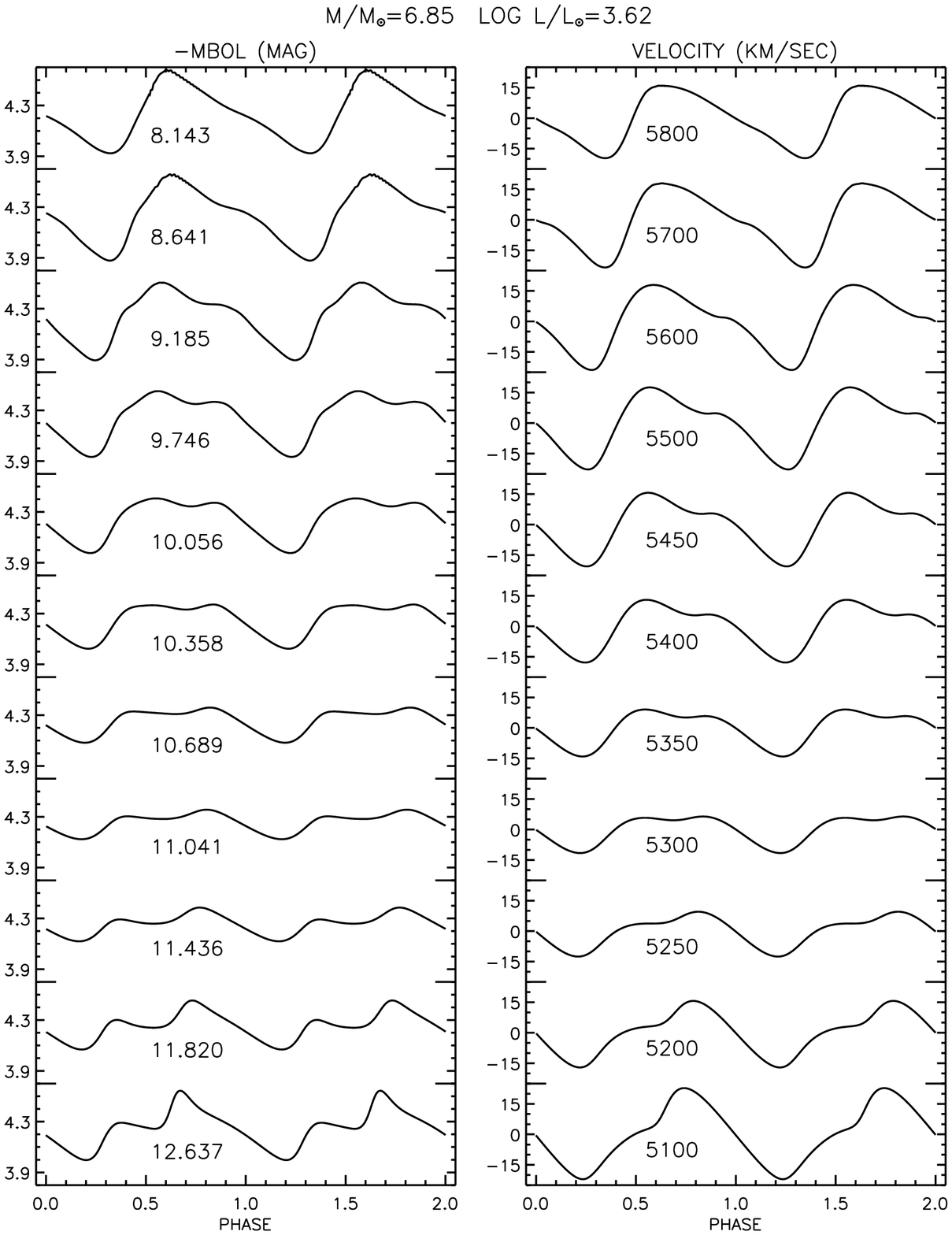,height=15cm,width=18cm} 
\caption {Similar to Fig. 1, but for \msun=6.85 models.} 
\end{figure*} 

\begin{figure*}[h] 
\psfig{figure=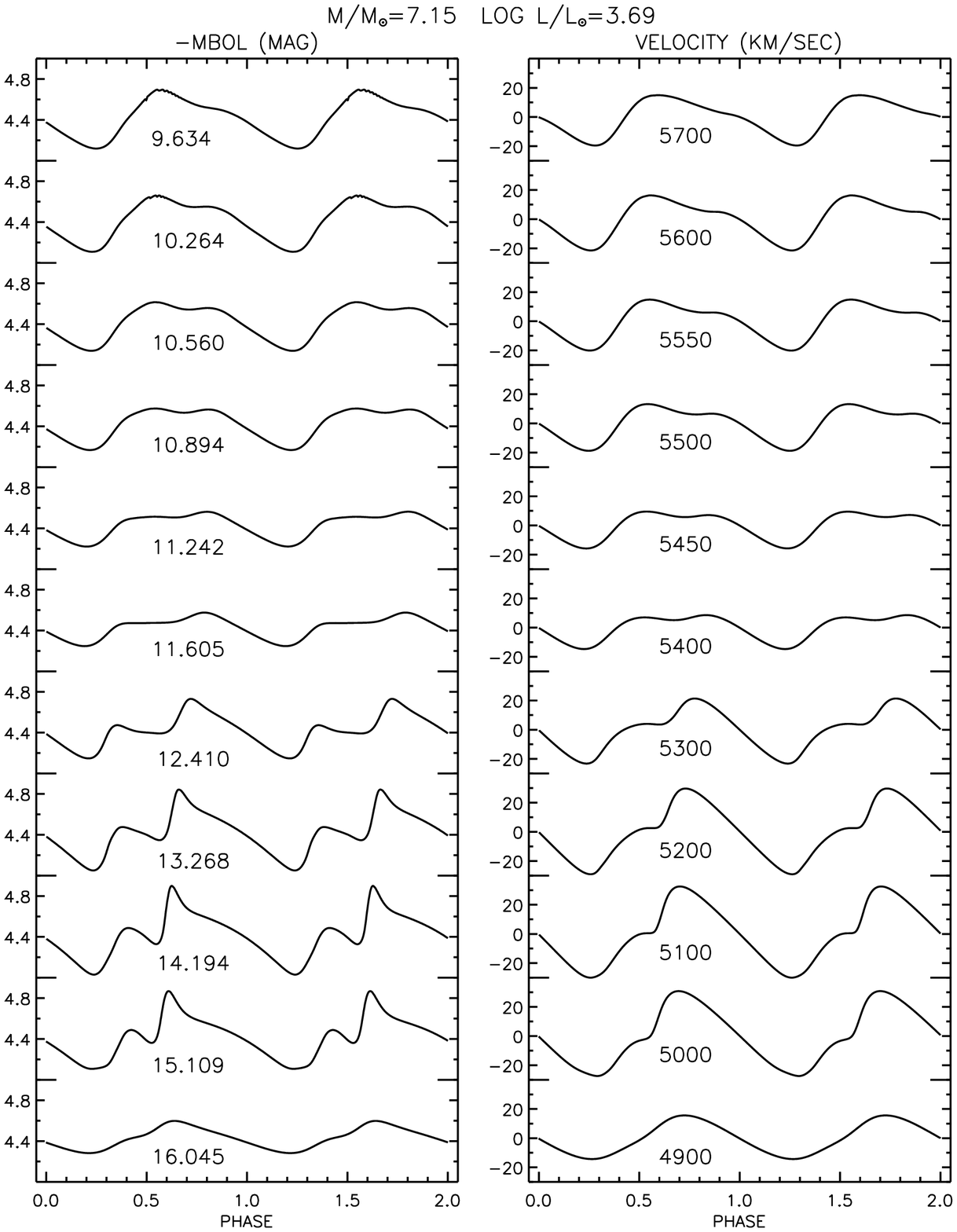,height=15cm,width=18cm} 
\caption {Similar to Fig. 1, but for a selection of models at \msun=7.15.} 
\end{figure*} 

Curves displayed in Figs. 1-6 show quite clearly that when 
moving from hotter to cooler effective temperatures the bump moves  
from the descending branch to the rising branch, i.e. an increase in 
the period moves the bump along the transition: descending branch 
$\rightarrow$ luminosity maximum $\rightarrow$ rising branch. 
Theoretical predictions in Figs. 1-6 disclose two key features:

a) the bump along the light curves crosses the luminosity maximum at 
shorter periods when compared with the velocity curves. This means 
that at fixed mass and for decreasing effective temperatures the Cepheid 
light curves start to show a flat-topped shape when the velocity curves 
still present the bump along the descending branch. This finding 
suggests that the lasting of the flat-topped phase in the luminosity 
curve is mainly governed by radius rather than by temperature variations. 

b) Along each sequence the model which attains the smallest velocity 
amplitude presents a flat-topped shape and the minimum in the luminosity 
amplitude coincides, within current temperature resolution, with the minimum 
in the velocity amplitude. On the basis of this evidence we will assume 
as period of the HP center -$P_{\rm{HP}}$- the period of the model which 
attains, along each sequence, the minimum amplitude in both luminosity  
and velocity changes.  

\begin{figure*}[h] 
\psfig{figure=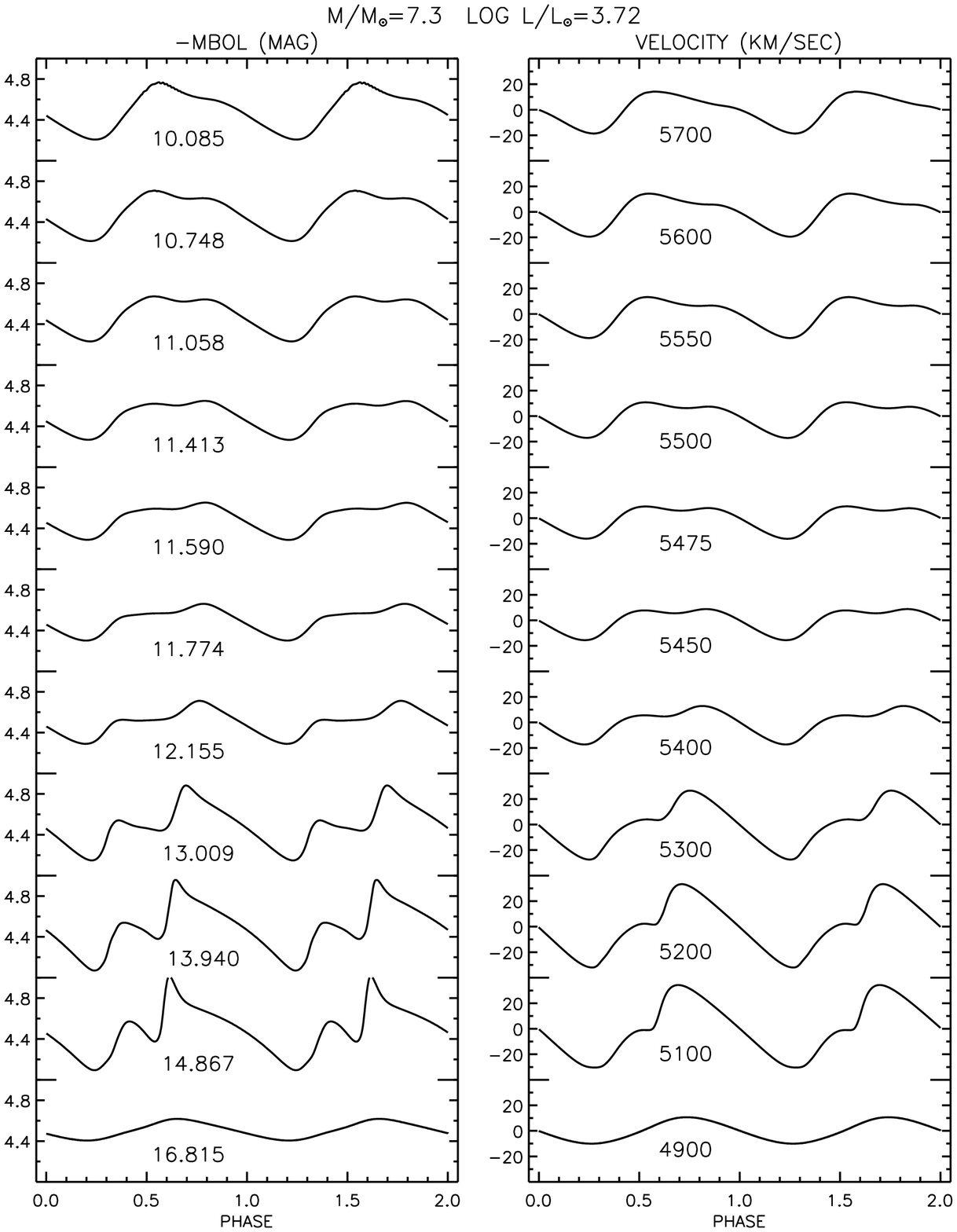,height=15cm,width=18cm} 
\caption {Similar to Fig. 1, but for a selection of models at \msun=7.3.} 
\end{figure*} 

\begin{figure*}[h] 
\psfig{figure=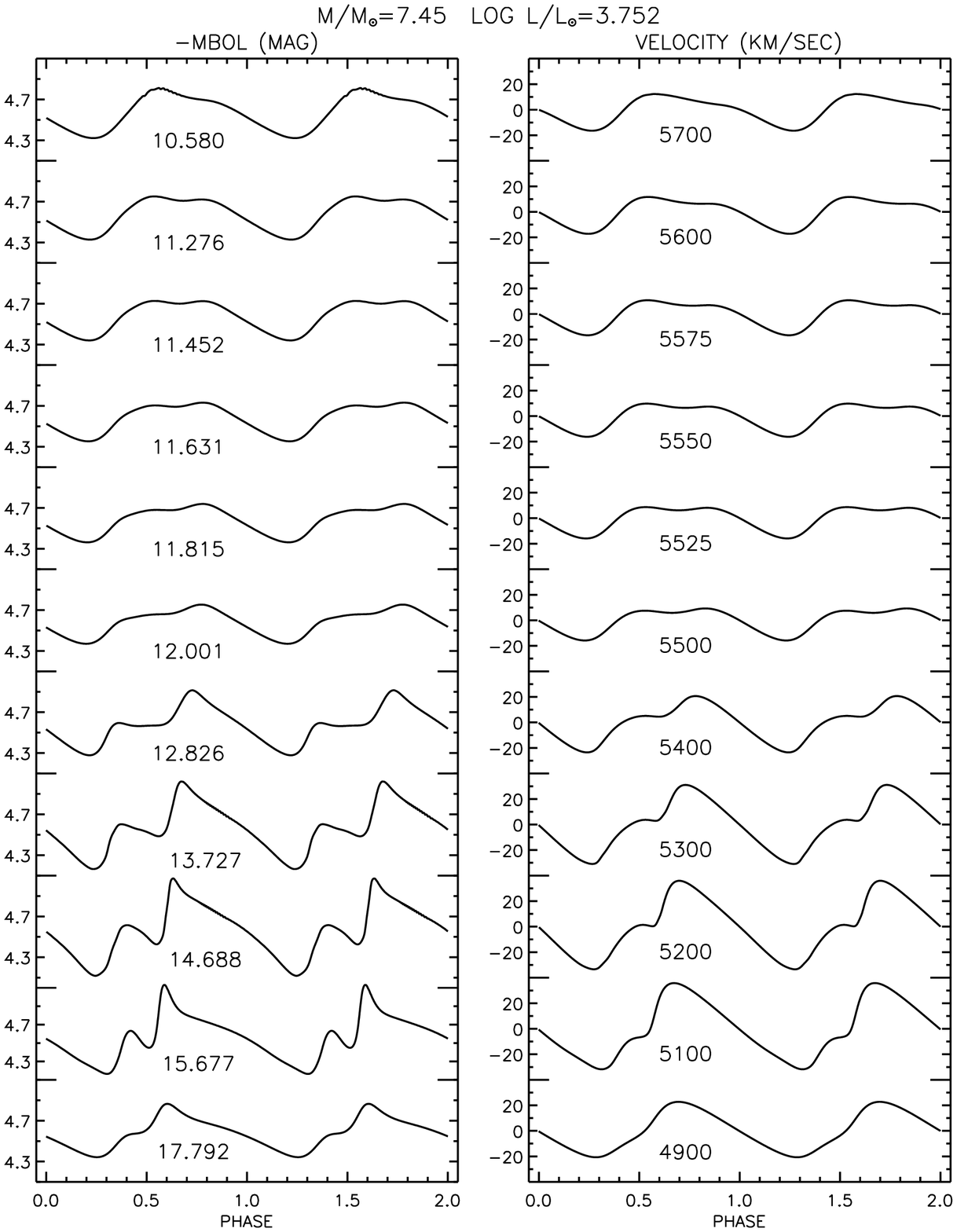,height=15cm,width=18cm} 
\caption {Similar to Fig. 1, but for a selection of models at \msun=7.45.} 
\end{figure*} 

\section{HP Systematic behavior}

In order to provide a quantitative estimate of the change in the 
pulsation properties we take into account three different 
parameters, namely the amplitude, the acuteness, and the skewness 
of both light and velocity curves. 
Although the Fourier parameters supply more detailed information 
on the dynamical behavior along the pulsation cycle, we adopted  
these three parameters because we are mainly interested in the HP 
center and also because they can be safely estimated even if the 
light and the velocity curves are not perfectly sampled. 

According to Stellingwerf \& Donohoe (1987) we define the acuteness 
as the ratio of the phase duration during which the magnitude is 
fainter than the median magnitude -$m_{\rm {med}}\,=\,0.5*(m_{\rm{max}}+m_{\rm {min}})$- 
to the phase duration during which it is brighter -$\phi_{\rm b}$- than 
$m_{\rm {med}}$, i.e. $Ac\, = (1/\phi_{\rm b})\, - 1$. 
This quantity is a measure of the top-down asymmetry of the light 
curve and it decreases when the shape changes from sawtooth to 
flat-topped. 
At the same time, we define the skewness as the ratio of the phase 
duration of the descending branch to the phase duration of the rising 
branch -$\phi_{\rm r}$-,  i.e. $Sk\, = (1/\phi_{\rm r})\, - 1$. 
This quantity is a measure of the asymmetry between rising and 
decreasing branch and it decreases when the slope of the rising branch 
becomes flatter. For symmetrical curves -not necessarily sinusoidal- 
both $Sk$ and $Ac$ attain values close to 1.  

Table 2 gives in the first two columns the mass and the effective 
temperature of each model, and in column 3) the ratio of second overtone 
to fundamental period. Since the second overtones are pulsationally 
stable in this region of the instability strip, we adopted for this 
mode the linear, nonadiabatic period. Columns 4) to 6) list the amplitude, 
the acuteness and the skewness of visual light curves. Bolometric magnitudes  
were transformed into V magnitudes by adopting bolometric corrections 
by Castelli et al. (1997). Columns 7) and 8) 
give the acuteness and the skewness of radial velocity curves.


Fig. 7 shows the luminosity amplitude in the V band -$A_{\rm V}$-  
(top panel), the skewness (middle panel), and the acuteness 
(bottom panel) as a function of the logarithmic period. Sequences 
characterized by different stellar masses (see labeled values) are
plotted using different symbols. Data plotted in Fig. 7 show that 
these three parameters typically present well-defined minima. We find 
that an increase in the stellar mass moves these minima toward longer
periods and that the periods of these minima are located at 
$P_{\rm{HP}}(A_{\rm V})=11.24\pm0.46$ d, $P_{\rm{HP}}(Ac)=11.17\pm0.48$ d, 
and $P_{\rm{HP}}(Sk)=10.73\pm0.97$ d, where the errors give the 
standard deviations. The period of the HP center based on $Sk$ minima 
is shorter than $P_{\rm{HP}}(A_{\rm V})$, and $P_{\rm{HP}}(Ac)$ because the sequence 
for \msun=6.55 does not show a sharp minimum.  

\begin{figure}[h] 
\psfig{figure=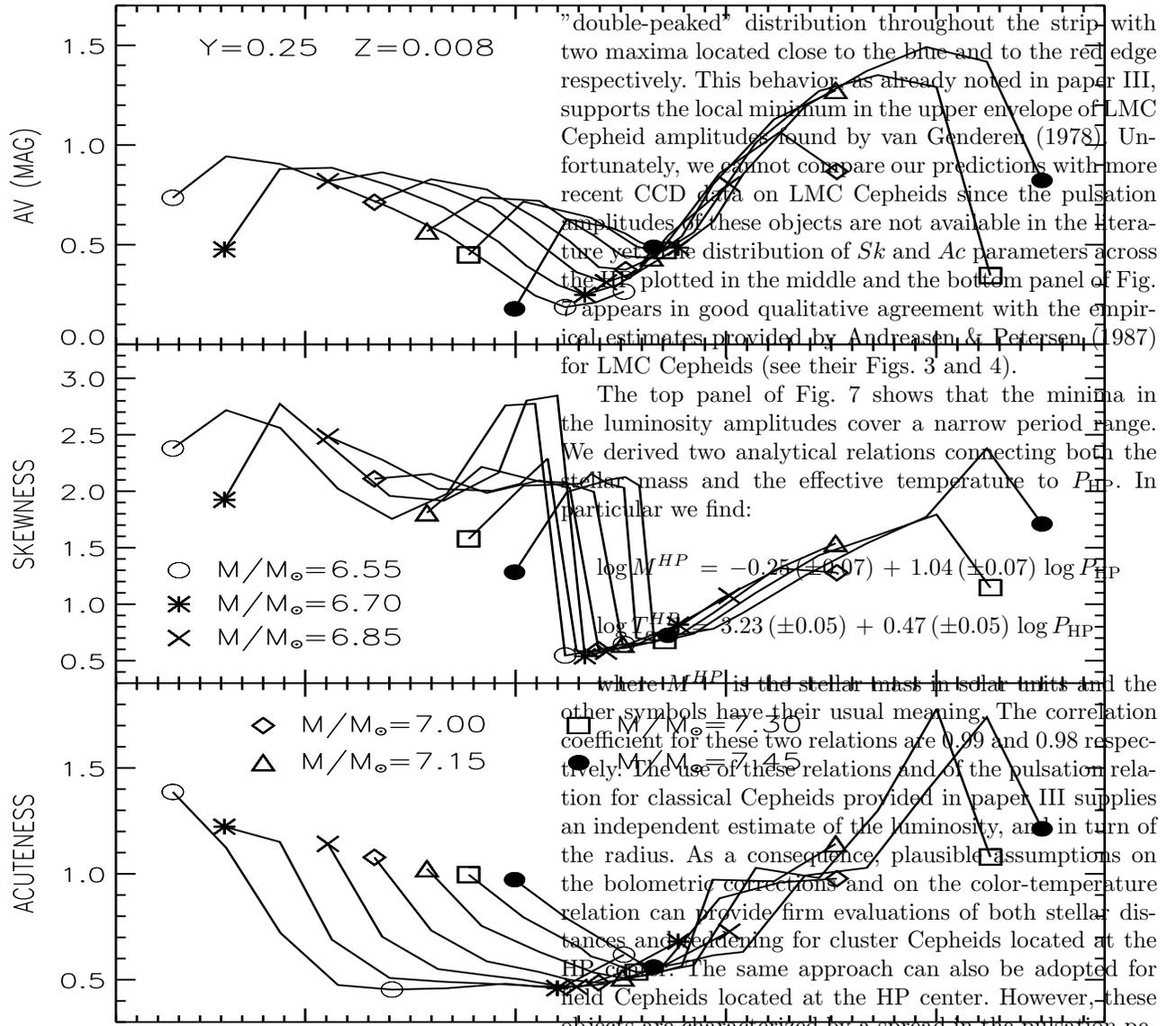,height=15cm,width=18cm} 
\caption {Luminosity amplitude in the V band (top panel), 
skewness (middle panel), and acuteness (bottom panel) as a function of 
logarithmic period. Models characterized by different stellar 
masses (see labeled values) were plotted using different symbols.} 
\end{figure} 

These theoretical minima appear in very good agreement with the 
empirical determination based on the Fourier parameters $\phi_{21}$ 
and $R_{21}$ of a large sample of LMC Cepheids provided by 
Welch et al. (1997) i.e. $P_{\rm{HP}}=11.2\;\pm0.8$ d and in reasonable 
agreement with the estimate, based on the same approach, provided by 
Beaulieu (1998) i.e. $P_{\rm{HP}}=10.5\;\pm0.5$ d. 
Here we note that the observed $P_{\rm{HP}}$ value was estimated as the 
period at which the Fourier parameters $\phi_{21}$ and $R_{21}$ of the 
light curves present a sudden jump, while the predicted $P_{\rm{HP}}$ values 
are the periods at which Bump Cepheid models attain the minimum value in 
$A_{\rm V}$, $Sk$, and $Ac$ respectively. However, Stellingwerf \& Donohoe (1986) 
using adiabatic, one-zone pulsation models showed that the jump in the 
$\phi_{21}$ parameter is correlated with a local minimum in the $Sk$ 
parameter. At the same time, Andreasen \& Petersen (1987) in a detailed 
analysis of both amplitudes and Fourier parameters found that the jump 
in $\phi_{21}$ and in $R_{21}$ of LMC Bump Cepheids takes place at the 
same period, within observational uncertainties, at which the B 
photographic amplitudes, the skewness, and the acuteness attain their 
minimum value. As a consequence, we conclude that not only $P_{\rm{HP}}(A_{\rm V})$
but also $P_{\rm{HP}}(Sk)$ and $P_{\rm{HP}}(Ac)$ are robust indicators of the HP 
center.  

Data plotted in the top panel show that when moving from the blue to 
the red edge the luminosity amplitudes do not show a monotonic behavior 
with stellar mass. In fact, close to the blue edge an increase in the 
stellar mass causes a decrease in the pulsation amplitudes, whereas 
close to the red edge the amplitudes present an opposite behavior. 
Moreover, the luminosity amplitudes present a "double-peaked" 
distribution throughout the strip with two maxima located close to 
the blue and to the red edge respectively. This behavior, as already 
noted in paper III, supports the local minimum in the upper envelope 
of LMC Cepheid amplitudes found by van Genderen (1978). Unfortunately, 
we cannot compare our predictions with more recent CCD data on LMC 
Cepheids since the pulsation amplitudes of these objects are not 
available in the literature yet.  
The distribution of $Sk$ and $Ac$ parameters across the HP plotted in 
the middle and the bottom panel of Fig. 7 appears in good qualitative 
agreement with the empirical estimates provided by Andreasen \& 
Petersen (1987) for LMC Cepheids (see their Figs. 3 and 4). 

The top panel of Fig. 7 shows that the minima in the luminosity amplitudes 
cover a narrow period range. We derived two analytical relations connecting 
both the stellar mass and the effective temperature to $P_{\rm{HP}}$. 
In particular we find:\\ 

$\log M^{HP}\,=\, -0.25\,(\pm0.07) \,+\,1.04\,(\pm0.07)\,\log P_{\rm{HP}}$\\ 

$\log T_{\rm e}^{HP}\,=\, 3.23\,(\pm0.05) \,+\, 0.47\,(\pm0.05)\,\log P_{\rm{HP}}$\\  

where $M^{HP}$ is the stellar mass in solar units and the other symbols 
have their usual meaning. The correlation coefficient for these two 
relations are 0.99 and 0.98 respectively. The use of 
these relations and of the pulsation relation for classical Cepheids 
provided in paper III supplies an independent estimate of the luminosity,  
and in turn of the radius. As a consequence, plausible assumptions on 
the bolometric corrections and on the color-temperature relation can 
provide firm evaluations of both stellar distances and reddening for 
cluster Cepheids located at the HP center.  
The same approach can also be adopted for field Cepheids located at 
the HP center. However, these objects are characterized by a spread 
in the pulsation period and therefore the masses and the effective 
temperatures derived using the previous relations can be affected 
by larger uncertainties. In fact, by adopting the empirical value 
-$P_{\rm{HP}}=11.2\pm0.8$ d- provided by Welch et al. (1997) we find 
$M^{HP}=7.0\pm1.0 M_{\odot}$, and $T_{\rm e}^{HP}=5300\pm600$ K. 

\begin{figure}[h] 
\psfig{figure=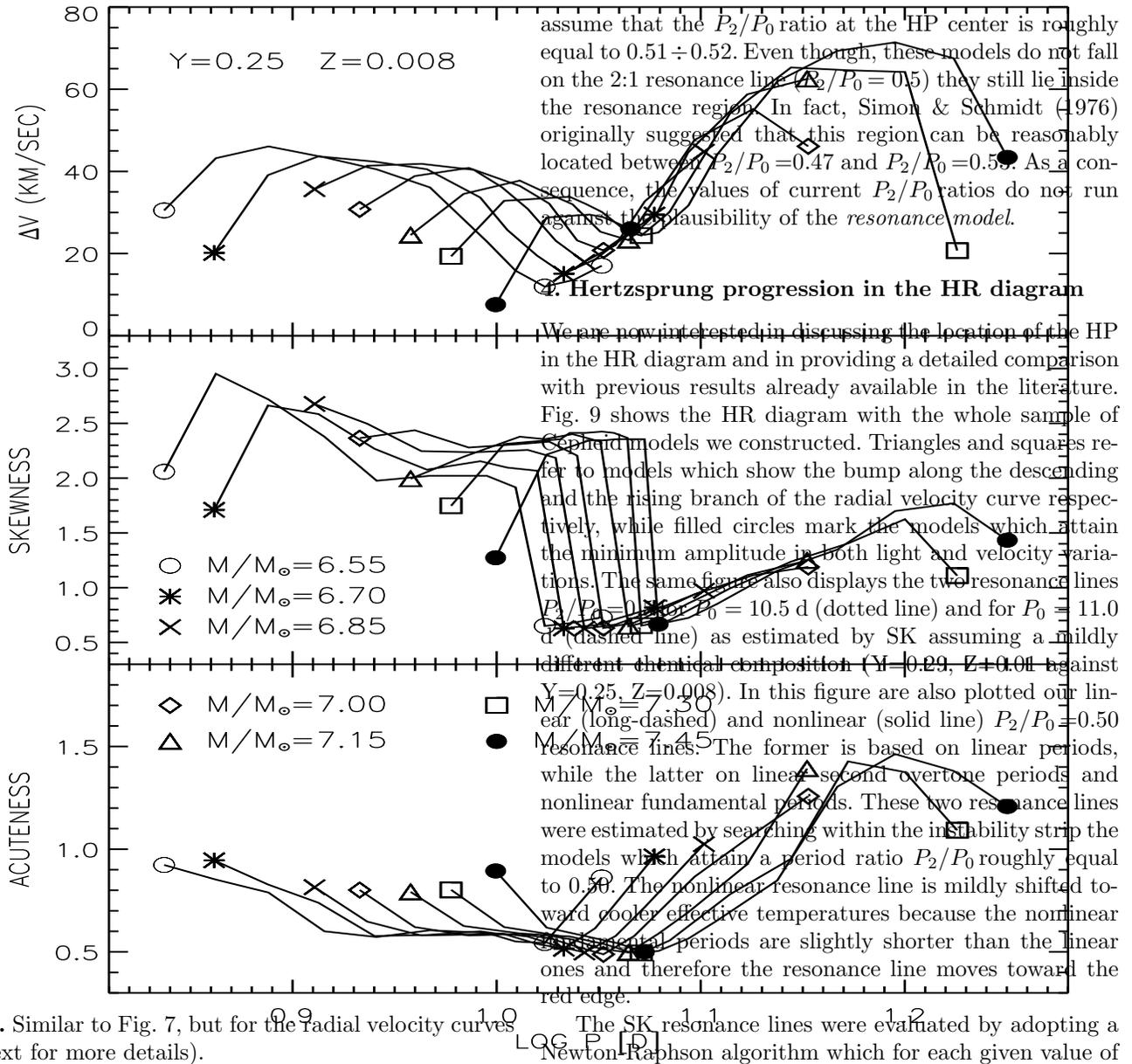,height=15cm,width=18cm} 
\caption {Similar to Fig. 7, but for the radial velocity curves 
(see text for more details).} 
\end{figure} 

Fig. 8 shows the same parameters of Fig. 7 but for the radial 
velocity curves. The behavior of $\Delta V$, $Sk$, and $Ac$ versus 
period are quite similar to the behaviors of the same parameters 
of the light curves. We find that the periods of the minima are now  
located at $P_{\rm{HP}}(\Delta V)=11.24\pm0.46$ d, $P_{\rm{HP}}(Ac)=11.29\pm0.53$ d, 
and $P_{\rm{HP}}(Sk)=11.27\pm0.49$ d respectively, and are once again in 
good agreement with empirical estimates.    
Although light and velocity curves show quite similar $Sk$ values over 
the whole period range, the velocity curves present larger $Sk$ values just 
before the HP center. This effect is due to the fact that the shape of 
the velocity curves becomes flat-topped at minimum amplitude whereas 
the light curves show the same shape at shorter periods (see \S 2). 
The comparison of predicted radial velocity amplitudes with observational 
data is difficult since spectroscopic measurement of LMC Cepheids are 
very scanty or do not cover the HP (Antonello 1998). However, radial 
velocity amplitudes for Galactic Cepheids (see Fig. 9 of paper III) 
collected by Cogan (1980) and by Bersier et al. (1994) show that 
the upper envelope presents a well-defined minimum 
across the HP center. The same outcome applies to the skewness, and 
indeed recent spectroscopic measurements of Galactic Cepheids 
(Gorynya 1998) show a sharp decrease close to $log P\approx1$. 
This provides a qualitative support to the predicted 
minima and to the "double-peaked" distributions shown in Fig. 8.  

Finally, we note that the \pve ratios (see Table 2) of the models 
located at the HP center range from 0.517 for the \msun=6.55 model 
to 0.512 for the \msun=7.45 model. These period ratios were estimated 
by adopting the linear second overtone periods, therefore we can safely 
assume that the \pve ratio at the HP center is roughly equal to 
$0.51\div0.52$. Even though, these models do not fall on the 2:1 resonance 
line (\pve$=0.5$) they still lie inside the resonance region.
In fact, Simon \& Schmidt (1976) originally suggested that this region 
can be reasonably located between \pve=0.47 and \pve=0.53. 
As a consequence, the values of current \pve ratios do not run against 
the plausibility of the {\em resonance model}.

\section{Hertzsprung progression in the HR diagram}

We are now interested in discussing the location of the HP in the 
HR diagram and in providing a detailed comparison with previous 
results already available in the literature.  
Fig. 9 shows the HR diagram with the whole sample of Cepheid 
models we constructed. Triangles and squares refer to models which show 
the bump along the descending and the rising branch of the radial velocity 
curve respectively, while filled circles mark the models which attain the 
minimum amplitude in both light and velocity variations. The same  
figure also displays the two resonance lines \pve=0.5 for $P_0=10.5$ d 
(dotted line) and for $P_0=11.0$ d (dashed line) as estimated by SK 
assuming a mildly different chemical composition (Y=0.29, Z=0.01 
against Y=0.25, Z=0.008). In this figure are also plotted our linear 
(long-dashed) and nonlinear (solid line) \pve=0.50 resonance lines.  
The former is based on linear periods, while the latter on linear 
second overtone periods and nonlinear fundamental periods. These 
two resonance lines were estimated by searching within the 
instability strip the models which attain a period ratio \pve 
roughly equal to 0.50. The nonlinear resonance line is mildly   
shifted toward cooler effective temperatures because the nonlinear 
fundamental periods are slightly shorter than the linear ones and 
therefore the resonance line moves toward the red edge.

\begin{figure}[h] 
\psfig{figure=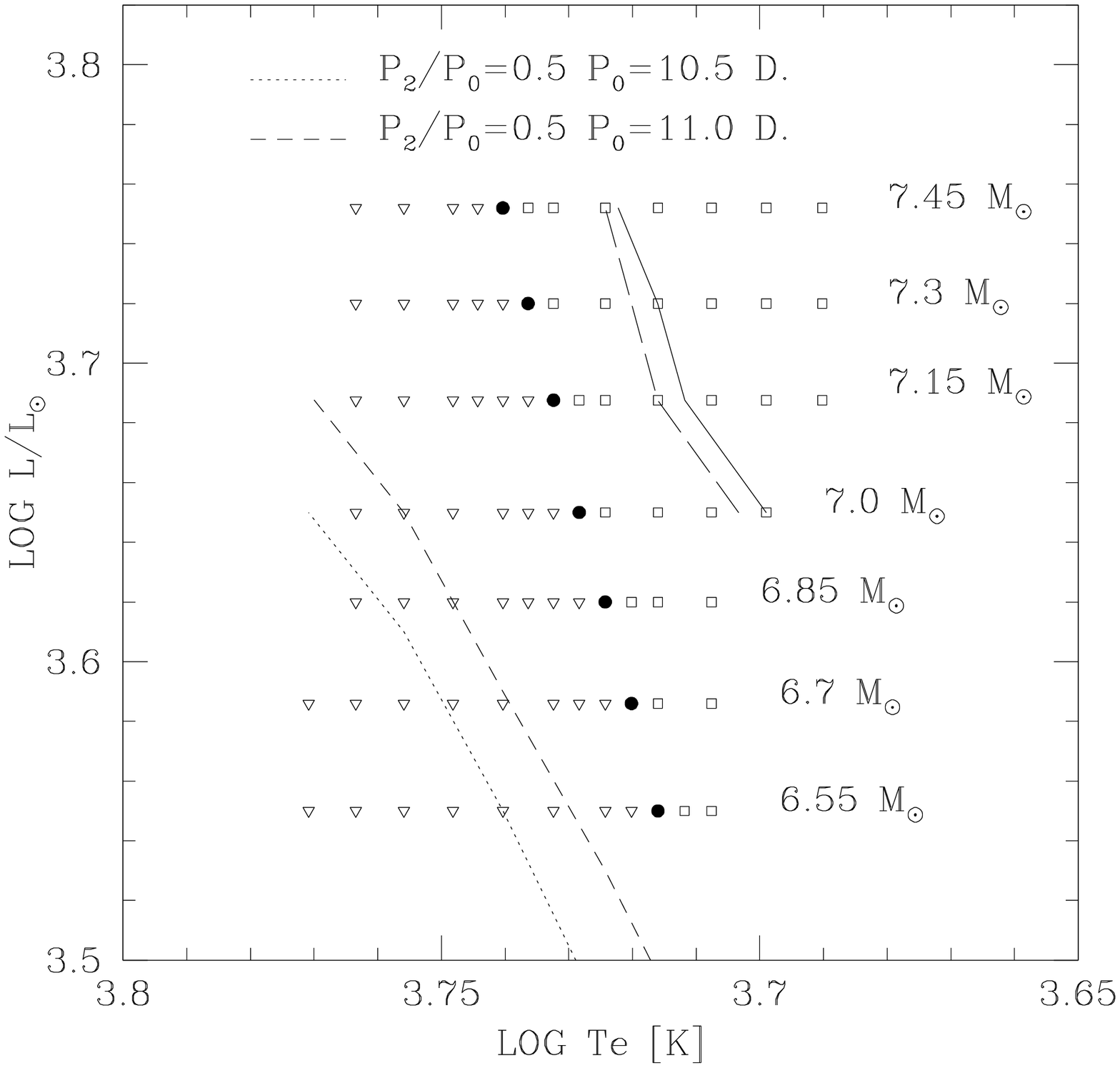,height=15cm,width=18cm} 
\caption {Distribution in the HR diagram of Bump Cepheid models. 
Cepheid models with a bump along the descending or the rising branch 
are plotted as triangles and squares. Cepheid models which attain 
along each sequence the minimum amplitude are plotted as filled 
circles. The dotted and the dashed lines refer to the linear \pve=0.5 
resonance lines derived by SK assuming that the resonance center 
is at $P_0=10.5$ d and 11.0 d respectively. The long-dashed and the 
solid lines show our linear and nonlinear \pve=0.50 resonance lines.} 
\end{figure} 

The SK resonance lines were evaluated by adopting a Newton-Raphson 
algorithm which for each given value of the effective temperature 
constructs a sequence of linear, nonadiabatic, radiative models by 
iterating on the values of mass and luminosity  until both the 
resonance condition -\pve=0.5- and the resonance center 
-$P_0=10.5, 11.0$ d- are matched. A slightly different approach 
for estimating the resonance lines in a linear regime was suggested 
by Buchler et al. (1996) but their predictions were not included 
in Fig. 9, because the effective temperature values were not 
provided. 

Fig. 9 shows that the SK resonance lines are located at hotter 
effective temperatures when compared with the nonlinear HP centers 
and this discrepancy increases toward higher luminosities. 
On the other hand, our linear and nonlinear resonance lines are 
located at effective temperatures systematically cooler that the 
nonlinear HP centers.  
The difference between our and SK resonance lines is not surprising, 
since our pulsation models are constructed by adopting different 
physical assumptions and chemical composition. However, 
we note that for each given luminosity the stellar masses adopted 
by SK are systematically smaller than our mass values and this 
difference typically increases when moving from lower to higher 
luminosities. 

The same difference appears in mass estimates, based on 
a similar approach, provided by Beaulieu \& Sasselov (1997, and 
references therein), and indeed for LMC Bump Cepheids at the resonance 
center $P_0=10.5$ d they give a mass value of 4.57 $M_\odot$. 
In order to test 
how the dynamical behavior depends on stellar mass, we constructed a 
new full amplitude, fundamental model by adopting the input parameters 
(\msun=5.35, \lsun=3.65, $T_{\rm e}$=5700 K) and chemical composition 
(Y=0.29, Z=0.01) used by SK.  
We selected this model since its location in the HR diagram is 
coincident with a model of our sequence for \msun=7.0 
(\lsun=3.65, $T_{\rm e}$=5700 K). Fig. 10 shows light (top) and velocity 
(bottom) curves  as a function of phase for our model (solid lines) 
and the model constructed by adopting the SK parameters (dashed lines). 
Although, the two models attain similar pulsation 
amplitudes (see Table 3) the shape of the curves is quite different. 
In fact, the bump in the less-massive model is located along the 
rising branch, while in the more-massive model along the descending 
branch. 

\begin{figure}[h] 
\psfig{figure=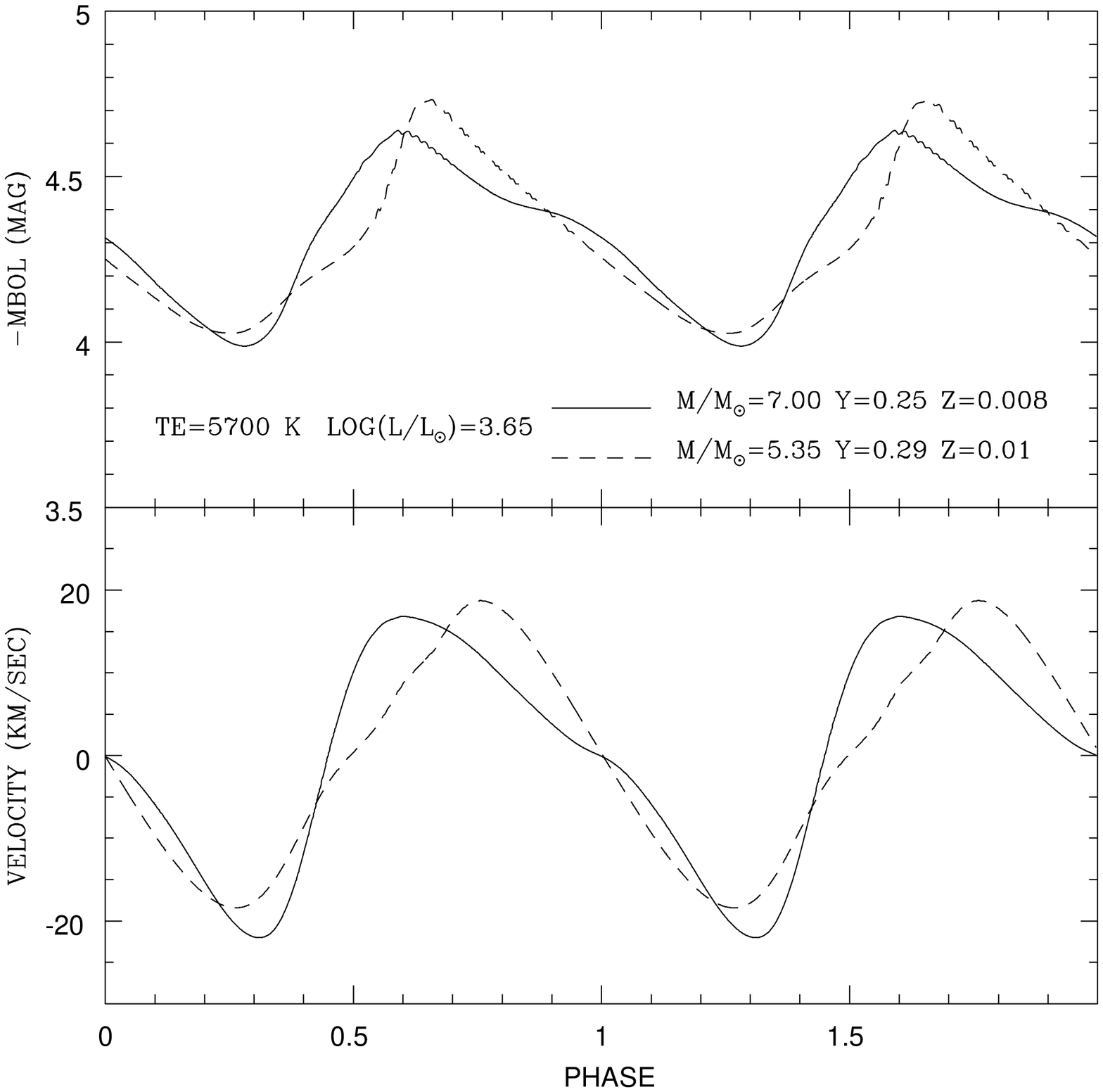,height=15cm,width=18cm} 
\caption {Light (top panel) and velocity (bottom panel) variations  
along two consecutive cycles. Solid and dashed lines refer to fundamental 
models constructed by adopting the same luminosity and effective 
temperature but different composition and mass value.}  
\end{figure} 

However, the key result disclosed by Fig. 10 is that the the
less-massive model, which is located very close to the 2:1 resonance
line (\pve $\approx0.494$, $P_0=11.0398$ d), presents almost symmetrical
curves and a small bump on the rising branch, whereas the curves of the
canonical model are more asymmetric around the maximum and show a
well-defined bump along the decreasing branch. This means that the
previous models should present quite different Fourier parameters. 
On the basis of these findings we can draw the following conclusion:
theoretical light and velocity curves of Bump Cepheid models depend 
on the adopted ML relation, therefore the comparison between theory 
and observations can supply independent constraints on this relation  
(Wood et al. 1997; paper I). Bump Cepheid models located at the HP 
center can provide tight constraints on theory, since the shape of light 
and velocity curves shows a stronger dependence on input parameters.  

In this context it is noteworthy that stellar masses based on 
linear period ratios are, as already noted by MBM, SK, and by 
Buchler et al. (1996), systematically smaller than the observed 
ones. On the other hand, our nonlinear, convective models 
based on a canonical ML relation agree with empirical Galactic 
Cepheid masses estimated by Gieren (1989) using the Baade-Wesselink 
method. In fact, by using the Gieren's Period-Mass relation 
(his relation 2) for $P_{\rm{HP}}=11.2$ d we obtain an empirical Cepheid mass 
of \msun$=6.9 \pm0.9$. This mass value is in satisfactory agreement,
within the observational uncertainties, with nonlinear predictions,
and indeed at the same period the theoretical masses 
range from 6.55 to 7.45 $M_\odot$. We know that we are comparing 
theoretical predictions for LMC Cepheids with observational 
data for Galactic Cepheids. However, current uncertainties on mass 
determinations are probably larger than the metallicity effect 
on Cepheid masses. In fact, preliminary theoretical results support 
the evidence that the Bump Cepheid masses at solar chemical composition 
are of the order of $6.5\pm0.25\, M_\odot$ (Ricci 1999).  
 
As a result, predictions based on nonlinear, convective models and on 
canonical evolutionary tracks settle down the long-standing conundrum 
of the Bump mass discrepancy. However a firm conclusion on this problem 
can be reached as soon as pulsation calculations constructed by adopting 
a noncanonical ML relation (i.e. based on evolutionary models which 
account for convective core overshooting) will become available. 
Note that accurate and homogeneous mass determinations based on 
the near-infrared surface brightness technique (Gieren et al. 1997) and on the infrared flux method (Fernley at al. 1989) 
can supply tight constraints on the Cepheid ML relation. 
Independent mass estimates can also be derived from orbits of 
Cepheids in binary systems. 
Interestingly enough, Evans et al. (1998) have recently observed 
with HST the hot binary companion of U Aql, a Galactic Bump Cepheid, 
and obtained for this variable a mass value of $5.1\pm0.7\, M_\odot$. 
Taken at face value this mass estimate is roughly 20\% smaller 
than the predicted mass range of Galactic Bump Cepheids 
previously mentioned. Unfortunately, up to now this is the only 
measurement of a Bump mass in a binary system and we still lack 
a firm evaluation of the error budget involved in these measurements 
(see their Table 2). In this context it is worth mentioning the 
direct measurements of Cepheid diameters through optical 
interferometry recently provided by Nordgren et al. (2000). 
As soon as new and more accurate interferometric data will become 
available, this approach can certainly supply an independent 
constraint on Bump Cepheid masses and in particular on the 
systematic uncertainties affecting empirical mass determinations.

\section{Dynamical behavior of Bump Cepheid models}

The appearance of the bump on the light curve is governed by 
temperature and radius variations. However, we focus our attention 
on radial velocity changes, since this is the observable generally 
adopted to investigate the intimate nature of the HP. In particular,
we discuss the dynamical behavior of two models located at the HP 
center and close to red edge, the systematics of the thermal structure 
will be addressed in a forthcoming paper (Bono et al. in preparation). 

\begin{figure*}[h]
\psfig{figure=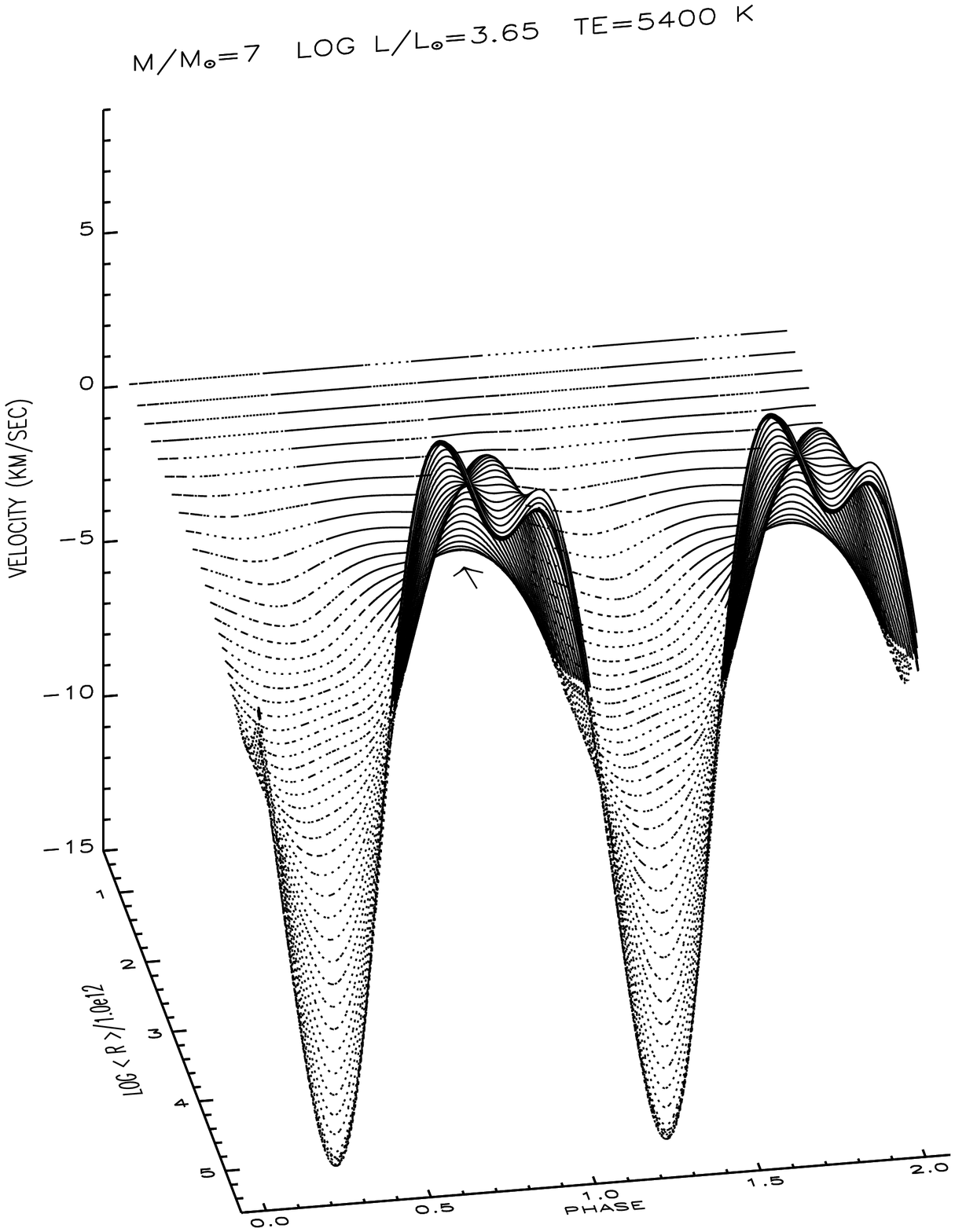,height=15cm,width=18cm} 
\caption {Radial velocity variations as a function of mean radius 
and phase for the model located close to the HP center, surface at bottom. 
Solid and dotted line refer to positive and negative velocities. The arrow 
marks the interaction between inner and outer layers.} 
\end{figure*} 

Fig. 11 shows in a 3D plot the 
velocity variations as a function of mean radius and phase
for a fundamental model located close to the HP center 
(\pve=0.516, $T_{\rm e}=5400$ K) of the sequence at \msun=7.0. 
Regions in which the radial velocity attains positive/negative 
values are plotted as solid/dotted lines. For avoiding misleading 
interpretations of the amplitude variations the radial velocities 
were plotted, according to Aikawa \& Whitney (1985), without 
applying any artificial shift or enhancement. 

Data plotted in Fig. 11 show that the outermost layers contract 
on a shorter time scale than the regions located just below them. 
In fact, the external layers reach the minimum 
contraction velocity at $\phi=0.23$ and then at $\phi=0.40$ 
-the phase of minimum radius- start to expand, whereas the underlying 
regions are still contracting. We measure phases with respect 
to the beginning of the surface contraction at maximum radius i.e. 
$\phi (R=R_{\rm{max}})=0$.  
Soon after this phase the outermost layers undergo at first a rapid 
expansion reaching the maximum expansion velocity at $\phi=0.54$ 
and then a decrease in the radial velocity due to gravity.   
During this slow down phase -$\phi\approx0.65$-  the outermost 
layers interact with deeper layers (see the arrow) that are 
rapidly moving outward and reach their maximum expansion velocity 
at $\phi\approx0.74$. The interaction between these two different 
dynamical behaviors causes an increase in the acceleration 
both toward the surface and toward the center. As a consequence 
in the outermost layers the radial velocity increases once again 
and then slows down showing a secondary maximum -the bump- 
at $\phi=0.85$. 
 
The innermost layers start to expand around $\phi=0.45$ but their 
expansion is not in phase with the outermost layers. In fact, the 
latter ones reach the maximum
expansion velocity at phases during which the bulk of the envelope
is still accelerating. This phase difference  causes the formation
of a shock between the slowing down motion of 
outermost layers and the outgoing expansion of deeper layers. 
As a consequence, the shocked region undergoes a strong compression 
which in turn generates pressure waves that propagate both toward 
the surface and toward the center of the stellar model. 
It is the outgoing pressure wave that causes, as already noted, 
the appearance of the bump, whereas the incoming pressure wave 
-the \cw- causes a short contraction phase in the innermost 
regions lasting from $\phi=0.7$ to $\phi=0.85$.  
Soon after the bounce at the stellar core at $\phi=0.95$, the \cw~  
at first moves rapidly outward and shortly delays the beginning of the 
contraction phase in the innermost regions and then limits the inward 
excursion of the layers crossed during its propagation out to the 
surface. This effect causes close to the phases of minimum radius 
a bounce between the overlying layers -that are rapidly contracting- 
and these layers that are contracting more slowly. 
The shock formed by this bounce generates two pressure waves:
the outgoing wave delays the outward excursion of the region 
located just below the outermost layers, while the incoming 
wave triggers the \cw.  
As a consequence, it is the propagation out to surface of the \cw~   
generated in the previous cycle that eventually causes the phase shift 
between the outermost layers and of the underlying regions, and in turn 
the appearance of the bump.    

A phase shift between the outermost layers and the bulk of the envelope 
was also noted by Karp (1975). However, he found that the \cw~ approaches 
the surface close to velocity maximum and not close to the bump phase, 
and therefore the {\em echo model} could not completely explain the HP. 
Our results support the timing suggested by Karp but it seems that 
it is the \cw~ which causes the phase shift, and in turn the bump.    
However, before a firm conclusion on the plausibility of this mechanism 
can be reached a detailed comparison between theoretical predictions and 
empirical data has to be provided.

\begin{figure*}[h] 
\psfig{figure=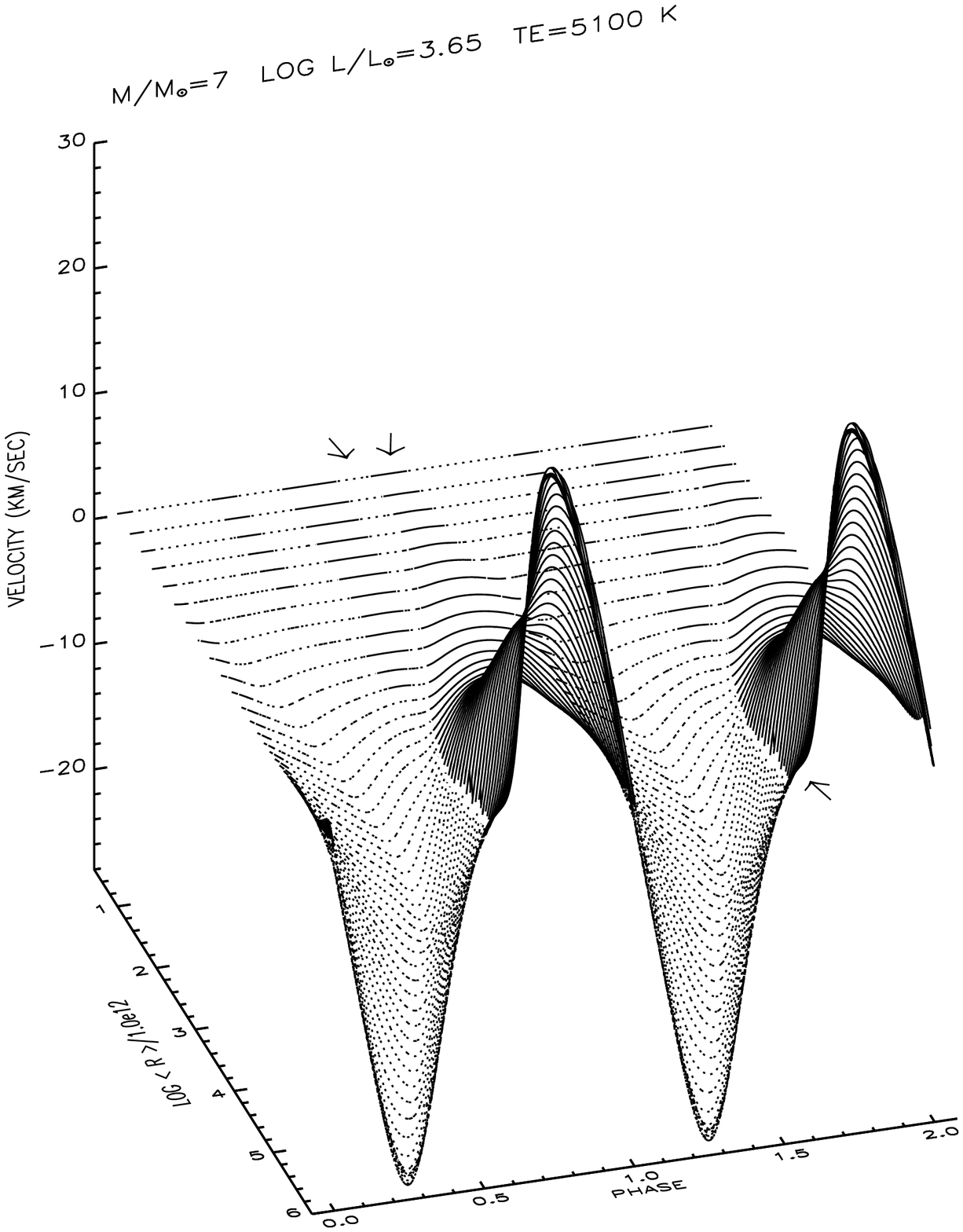,height=15cm,width=18cm} 
\caption {Radial velocity variations as a function of mean radius
and phase for a model located close to the red edge, surface at bottom.
The two arrows at the base of the envelope mark the approach and the 
ensuing reflection of the \cw~ at the stellar core, while the arrow at 
the surface marks the phases at which appears the shoulder.} 
\end{figure*}

We now briefly discuss the dynamical behavior of a model located 
close to the red edge (\pve=0.502, $T_{\rm e}=5100$ K) of the sequence 
at \msun =7.0, since it presents an interesting feature worth 
being investigated.  Fig. 12 shows the radial velocity 
as viewed from the stellar surface. 
The overall behavior is similar to the model located close to the 
HP center, but in this red model the formation and the incoming 
propagation of the \cw~ can be more easily identified. In fact, 
close to $\phi\approx 0.5$ and $ \log {<R> / 10^{12}}\approx3.5$  
its appearance is marked by a narrow region in which the 
velocity attains negative values and then moves inward reaching 
the base of the envelope at $\phi\approx 0.8$ (see the arrows).   
In this model the contraction of the overlying layers is more rapid 
than for the model close to the HP center, and indeed the \cw~  
stops the expansion of the underlying layers. This fact and  
the evidence that the outermost layers start to expand at later 
phases causes the appearance along the surface velocity curve 
of a shoulder close to $\phi=0.5\div0.55$ (see the arrow at the 
top of the envelope). Soon after these 
layers experience a rapid outgoing acceleration and this 
motion is almost in phase with the motion of the envelope 
and therefore the outermost layers do not 
interfere with the underlying regions since they are almost 
co-moving. It turns out that in this model the bump, i.e. what 
we defined as the secondary maximum, is the main maximum while 
the true maximum is the bump which appears on the rising branch. 

This finding provides a straightforward explanation to what has 
been defined by FSS as {\em a happy but also ill-understood 
circumstance}. In fact, these authors in order to derive a linear 
relationship between period, phase of the bump and Bump Cepheid 
radii noted that from an observational point of view it could 
not be firmly stated whether the primary or the secondary maximum 
was generated by the \cw. To overcome this thorny problem
they suggested to measure the phase of the secondary maximum if the 
bump is located along the descending branch and the phase of the 
primary maximum if the bump appears along the rising branch.  
The dynamical behavior of the Bump Cepheid model located close to the 
red edge supplies a plain theoretical support to this far-sighted 
observational choice. At the same time, current models also explain 
why light and velocity amplitudes show a "double-peaked" distribution 
when moving from the blue to the red edge of the instability strip. 
The maximum located close to the blue edge is typical of 
relatively short-period Cepheids (see paper III), whereas the 
second one is caused by the HP.  

\section{Summary and conclusions}

This paper presents the results of an extensive theoretical investigation on 
Bump Cepheids. In order to provide a detailed analysis of the pulsation 
behavior of these objects we computed  several sequences of full amplitude, 
nonlinear, convective models at fixed chemical composition (Y=0.25, Z=0.008). 
The models were constructed by adopting a mass step of 0.15$M_\odot$ and a 
temperature step of 100 K close to the instability edges and of 50 K close 
to the HP center. The main outcomes of this analysis are the following:

1) theoretical light and velocity curves account for the HP, and indeed 
close to the blue edge the bump is located along the descending branch, 
at longer periods it crosses at first the luminosity/velocity maximum 
and then appears along the rising branch. 

2) In a very narrow period range both light and velocity curves 
show a flat-topped shape and their amplitudes attain a well-defined 
minimum. The predicted period of the minima -$P_{\rm{HP}}=11.24\pm0.46$ d- 
is in very good agreement with the empirical value found by 
Welch et al. (1997) for LMC Cepheids i.e. $P_{\rm{HP}}=11.2\pm0.8$ d. 
Moreover, light and velocity amplitudes present a "double-peaked" 
distribution which agrees with observational data. 
 
3) Both the skewness and the acuteness of light and velocity curves show 
a well-defined minimum at the HP center and their distributions are in 
good qualitative agreement with empirical estimates provided by 
Andreasen \& Petersen (1987). The periods of the HP center are, within 
the uncertainties, in good agreement with observational values.  
  
4) {\em The models at the HP center are located within the 
resonance region but not on the 2:1 resonance line} (\pve=0.5), 
and indeed the \pve ratios range from 0.51 to 0.52.  

5) Predicted Bump Cepheid masses, based on a ML relation which neglects
the convective core overshooting, are in good agreement with the empirical 
masses of Galactic Cepheids estimated by adopting the Baade-Wesselink 
method (Gieren 1989). In fact, the observed mass range at the HP center 
-$P=11.2$ d- is $6.9\pm0.9\; M_\odot$, while the theoretical 
one is $7.0\pm0.45\; M_\odot$. Even if Galactic Cepheids are more 
metal-rich than LMC Cepheids, this result seems to settle down the 
long-standing problem of the Bump mass discrepancy.  
 
The results presented in this paper were mainly aimed at testing whether 
current hydrodynamical models which include the coupling between pulsation 
and convection together with a ML relation based on canonical evolutionary 
models account for the HP in LMC Cepheids. It is clear that the 
theoretical scenario we developed seems to provide a reliable description
of several empirical facts. Moreover, we confirm that Bump Cepheid models 
can supply fundamental constraints on stellar masses, and in turn on the ML 
relation of intermediate-mass stars. We also note that current Cepheid 
metal-intermediate, convective models do show plausible pulsation 
properties. 

Two crucial topics need to be properly addressed before firm conclusions
on the HP can be reached. To validate present nonlinear Cepheid models 
a thorough theoretical analysis of the Fourier parameters should be 
provided together with a detailed comparison with observational data 
available in the literature (Bono et al. in preparation). At the same time, the 
dependence of the HP behavior both on the chemical composition should 
be investigated as well. 
New and accurate CCD data on light and velocity amplitudes as well as 
on the skewness and the acuteness of these curves can certainly improve 
the location of the HP center and feed the future investigations with 
robust observational constraints. 

\begin{acknowledgements} 
It is a pleasure to thank V. Castellani and F. Caputo for a detailed 
reading of an early draft of this paper and for many interesting 
and enlightening discussions on Cepheids. We wish also to acknowledge 
N. Remage Evans and T. Nordgren for insightful discussions on current 
empirical estimates of Cepheid masses. We also acknowledge an anonymous
referee for some useful suggestions that improved the readability
of the paper.     
\end{acknowledgements}


\end{document}


\small 
\begin{flushleft} 
\begin{table}  
\begin{tabular}{ccccccccccccc}
\multicolumn{13}{c}{Table 1: Nonlinear Observables}\\  
\noalign{\smallskip}\noalign{\smallskip}\hline\noalign{\smallskip}  
$M^a$&
$\log L^b$&
$T_{\rm e}^c$&
$P_0^d$& 
$\log{\overline{R}}^e$& 
$\Delta R/R_{\rm {ph}}^f$& 
$\Delta u^g$&
$\Delta M_{\rm {bol}}^h$& 
$\Delta \log g_{\rm s}^i$&  
$\Delta \log g_{\rm e}^j$&
$\Delta T^k$&
$\Delta T_{\rm e}^l$&
$KE^m$\\
(1)&
(2)&
(3)&
(4)&
(5)&
(6)&
(7)&
(8)& 
(9)& 
(10)& 
(11)&
(12)&
(13)\\  
\hline\noalign{\smallskip} 
6.55& 3.550 & 5900 & 6.8718 & 1.758 & 0.072 & 30.60 &0.601 &0.06 &0.35 & 700& 900& 9.1320(40) \\
6.55& 3.550 & 5800 & 7.2849 &1.773 & 0.098 & 43.18 &0.758 &0.09 &0.48 & 900& 1100& 1.5219(41)\\
6.55& 3.550 & 5700 & 7.7306 &1.789 &0.102  & 46.07  & 0.714 & 0.09  & 0.51 &800 &1000 &1.8923(41)\\
6.55& 3.550 & 5600 & 8.2336 &1.803 & 0.096 & 43.41 &0.615 &0.08 &0.51 & 650& 800& 1.7991(41)\\
6.55& 3.550 & 5500 & 8.7338 &1.818 & 0.089 & 40.47 &0.516 &0.08 &0.47 & 550& 700& 1.8008(41)\\
6.55& 3.550 & 5400 & 9.2755 &1.833 & 0.083 & 36.08 &0.402 &0.07 &0.42 & 500& 600& 2.3552(41)\\
6.55& 3.550 & 5300 & 9.8892 &1.847 & 0.058 & 22.94 &0.243 &0.05 &0.29 & 300& 400& 1.0706(41)\\
6.55& 3.550 & 5250 & 10.2164 &1.854 & 0.043 & 15.81 &0.169 &0.04 &0.21 & 250& 300& 5.1556(40)\\
6.55& 3.550 & 5200 & 10.5579 &1.861 & 0.033 & 11.95 &0.136 &0.03 &0.16 & 150& 200& 1.5752(40)\\
6.55& 3.550 & 5150 & 10.9123  & 1.869 & 0.036 & 13.59 &0.146 &0.03 &0.16 &150 &200 & 2.9776(39)\\
6.55& 3.550 & 5100 & 11.2589 &1.877 & 0.045 & 17.03 &0.179 &0.04 &0.19 & 200& 250& 5.0499(39)\\
6.70& 3.586 & 5900 & 7.2703 & 1.775& 0.048 & 20.16 &0.385 &0.04 &0.22 & 450& 600& 4.4741(40)\\
6.70& 3.586 & 5800 & 7.7257 &1.791 & 0.090 & 39.08 &0.704 &0.08 &0.43 & 850& 1000& 1.4157(41)\\
6.70& 3.586 & 5700 & 8.1787 &1.806 & 0.098 & 43.64 &0.697 &0.09 &0.49 & 800& 950& 1.8219(41)\\
6.70& 3.586 & 5600 & 8.7118 &1.821 & 0.094 &42.44  & 0.616 & 0.08 &0.50 &700 &850 & 3.4467(40)\\
6.70& 3.586 & 5500 & 9.2427 &1.836 & 0.091 &40.51  & 0.516 &0.08 &0.47 &600 &700 & 2.5989(41)\\
6.70& 3.586 & 5400 & 9.8168 &1.855 & 0.080 & 33.62 & 0.372 &0.07 &0.41 &450 &550 & 2.3430(41)\\
6.70& 3.586 & 5350 & 10.1217 &1.857 & 0.065 & 25.74 & 0.277 &0.06 &0.33 &350 &450 & 1.4305(41)\\
6.70& 3.586 & 5300 & 10.4676 &1.865 & 0.052 & 19.50 & 0.222 &0.05 &0.27 &300 &350 & 8.7687(40)\\
6.70& 3.586 & 5250 & 10.7876 &1.872 & 0.042 & 15.07 & 0.183 &0.04 &0.22 &200 &250 & 3.4250(40)\\
6.70& 3.586 & 5200 & 11.1770 &1.880 & 0.047 & 18.62 & 0.212 &0.04 &0.23 &250 &300 & 1.1107(40)\\
6.70& 3.586 & 5100 & 11.9479 &1.896 & 0.076 & 29.47 & 0.327 &0.07 &0.32 &400 &500 & 2.5620(40)\\
6.85& 3.620 & 5800 & 8.1429 &1.808 & 0.084 & 35.63 & 0.656 &0.07 &0.39 &750 &950 & 1.2515(41)\\
6.85& 3.620 & 5700 & 8.6411 &1.823 & 0.094 & 41.36 & 0.679 &0.08 &0.48 &750 &950 & 1.2991(41)\\
6.85& 3.620 & 5600 & 9.1850 &1.838 & 0.094 & 41.82 & 0.611 &0.08 &0.49 &700 &850 & 1.8822(41)\\
6.85& 3.620 & 5500 & 9.7459 &1.853 & 0.093 & 40.36 & 0.517 &0.08 &0.48 &600 &750 & 3.1404(41)\\
6.85& 3.620 & 5450 & 10.0558 &1.860 & 0.086 & 36.20 & 0.430 &0.08 &0.44 &500 &650 & 2.7778(41)\\
6.85& 3.620 & 5400 & 10.3578 &1.867 & 0.076 & 30.85 & 0.347 &0.07 &0.40 &450 &550 & 2.2254(41)\\
6.85& 3.620 & 5350 & 10.6894 &1.875 & 0.061 & 23.11 & 0.276 &0.05 &0.32 &350 &400 & 1.3148(41)\\
6.85& 3.620 & 5300 & 11.0408 &1.882 & 0.052 & 17.97 & 0.232 &0.04 &0.27 &250 &300 & 6.0248(40)\\
6.85& 3.620 & 5250 & 11.4360 &1.889 & 0.055 & 22.05 & 0.265 &0.05 &0.29 &300 &350 & 2.2124(40)\\
6.85& 3.620 & 5200 & 11.8202 &1.898 & 0.077 & 32.79 & 0.385 &0.07 &0.39 &450 &550 & 2.3564(40)\\
6.85& 3.620 & 5100 & 12.6365 &1.916 & 0.112 & 44.79 & 0.544 &0.10 &0.53 &650 &800 & 7.7163(40)\\
7.15& 3.690 & 5800 & 9.0744 &1.842 & 0.060  & 24.48  &0.453 &0.05 &0.27 &550 &650 & 3.9501(40)\\
7.15& 3.690 & 5700 & 9.6344 &1.858 & 0.081 & 34.50 &0.576 &0.07 &0.41 &700 &850 & 5.9840(40)\\
7.15& 3.690 & 5600 & 10.2641 &1.873 & 0.090 & 37.74 &0.550 &0.08 &0.46 &650 &800 & 2.4011(41)\\
7.15& 3.690 & 5550 & 10.5597 &1.880 & 0.085 & 34.96 &0.476 &0.07 &0.44 &550 &700 & 2.3934(41)\\
7.15& 3.690 & 5500 & 10.8941 &1.887 & 0.081 & 31.94 &0.407 &0.07 &0.43 &500 &600 & 2.2713(41) \\
7.15& 3.690 & 5450 & 11.2419 &1.895 & 0.068 & 25.25 &0.340 &0.06 &0.36 &350 &450 & 1.4116(41)  \\
7.15& 3.690 & 5400 & 11.6051 &1.902 & 0.063 & 23.24 &0.327 &0.05 &0.34 &350 &400 & 8.6794(40)  \\
7.15& 3.690 & 5350 & 11.9971 &1.910 & 0.074 & 30.24 &0.400 &0.06 &0.39 &450 &550 & 3.7647(40) \\
7.15& 3.690 & 5300 & 12.4099 &1.919 & 0.103 & 44.55 &0.581 &0.09 &0.53 &650 &800 & 7.2956(40)  \\
7.15& 3.690 & 5200 & 13.2676 &1.938 & 0.142 & 58.72 &0.789 &0.12 &0.74 &900 &1150 &1.7414(41)  \\
7.15& 3.690 & 5100 & 14.1940 &1.955 & 0.163 & 62.48 &0.865 &0.14 &0.83 &1050 &1250 & 2.4528(41) \\
7.15& 3.690 & 5000 & 15.1090 &1.970 & 0.166 & 58.15 &0.758 &0.15 &0.77 &950 &1150 & 3.2418(41)\\
7.15& 3.690 & 4900 & 16.0454 &1.980 & 0.094 & 29.91 &0.314 &0.08 &0.35 &400 &500 & 1.0253(41) \\
\hline\noalign{\smallskip}
\end{tabular}
\end{table}
\end{flushleft}

\begin{flushleft} 
\begin{table}  
\begin{tabular}{ccccccccccccc}
\multicolumn{13}{c}{Table 1: Continued}\\  
\noalign{\smallskip}\noalign{\smallskip}\hline\noalign{\smallskip}  
$M^a$&
$\log L^b$&
$T_{\rm e}^c$&
$P_0^d$& 
$\log{\overline{R}}^e$& 
$\Delta R/R_{\rm {ph}}^f$& 
$\Delta u^g$&
$\Delta M_{\rm {bol}}^h$& 
$\Delta \log g_{\rm s}^i$&  
$\Delta \log g_{\rm e}^j$&
$\Delta T^k$&
$\Delta T_{\rm e}^l$&
$KE^m$\\
(1)&
(2)&
(3)&
(4)&
(5)&
(6)&
(7)&
(8)& 
(9)& 
(10)& 
(11)&
(12)&
(13)\\  
\hline\noalign{\smallskip} 
7.30& 3.720 & 5800 & 9.4996 &1.857 & 0.048 &19.32 &0.358 &0.04 &0.21 &450 &550  & 2.5048(40)\\
7.30& 3.720 & 5700 & 10.0848 &1.872 & 0.071 &32.81 &0.560 &0.07 &0.40 &650 &800  &8.2223(40) \\
7.30& 3.720 & 5600 &10.7479  &1.888 & 0.083 &33.67 &0.493 &0.07 &0.43 &600 &700  &1.9504(41) \\
7.30& 3.720 & 5575 &10.8600  &1.891 & 0.084 &33.78 &0.480 &0.07 &0.44 &550 &700  &2.1880(41) \\
7.30& 3.720 & 5550 &11.0578  &1.895 & 0.081 &32.06 &0.440 &0.07 &0.43 &500 &650  &2.1052(41)\\
7.30& 3.720 & 5525 &11.2599  &1.899 & 0.078 &30.04 &0.397 &0.07 &0.41 &500 &600  &1.9245(41)\\
7.30& 3.720 & 5500 &11.4131  &1.903 & 0.074 &27.89 &0.380 &0.06 &0.40 &450 &550 & 1.6865(41)\\
7.30& 3.720 & 5475 &11.5898  &1.906 & 0.070 &25.31 &0.363 &0.06 &0.38 &400 &450 & 1.3366(41)\\
7.30& 3.720 & 5450 & 11.7741 &1.910 & 0.067 &24.35 &0.359 &0.06 &0.36 &350 &450 & 9.9195(40)\\
7.30& 3.720 & 5400 & 12.1549 &1.918 & 0.075 &30.11 &0.420 &0.07 &0.40 &450 &550 & 5.4813(40)\\
7.30& 3.720 & 5300 & 13.0085 &1.936 & 0.126 &53.97 &0.731 &0.11 &0.65 &800 &1000 &1.3798(41)\\
7.30& 3.720 & 5200 & 13.9396 &1.955 & 0.160 &65.25 &0.885 &0.14 &0.84 &1050 &1250 &2.5421(41)\\
7.30& 3.720 & 5100 & 14.8666 &1.971 & 0.176 &64.60 &0.917 &0.15 &0.89 &1100 &1350 &3.3421(41) \\
7.30& 3.720 & 5000 & 15.8518 &1.987 & 0.184 &64.19 &0.879 &0.16 &0.87 &1100 &1350 &5.2853(41) \\
7.30& 3.720 & 4900 & 16.8146 &1.994 & 0.067 &20.71 &0.212 &0.06 &0.24 &300 &350 &4.9524(40) \\
7.45& 3.752 & 5800 & 9.9895 &1.872 & 0.018  & 7.58  &0.142 &0.02 &0.08 &150 &200 &7.0975(39)\\
7.45& 3.752 & 5700 & 10.5798 &1.888 & 0.071 & 28.74 &0.489 &0.06 &0.36 &600 &700 &6.9298(40)\\
7.45& 3.752 & 5625 & 11.1041 &1.900 & 0.075 & 29.46 &0.446 &0.07 &0.39 &550 &650 &1.3650(41)\\
7.45& 3.752 & 5600 & 11.2759 &1.903 & 0.075 & 28.84 &0.422 &0.06 &0.39 &500 &650 &1.4235(41)\\
7.45& 3.752 & 5575 & 11.4520 &1.907 & 0.073 & 27.43 &0.386 &0.06 &0.38 &450 &600 &1.3788(41)\\
7.45& 3.752 & 5550 & 11.6314 &1.911 & 0.071 & 25.98 &0.378 &0.06 &0.37 &450 &550 &1.2519(41)\\
7.45& 3.752 & 5525 & 11.8146 &1.914 & 0.069 & 24.52 &0.375 &0.06 &0.37 &400 &500 &1.0727(41)\\
7.45& 3.752 & 5500 & 12.0010 &1.919 & 0.069 & 25.09 &0.383 &0.06 &0.37 &400 &500 &8.3532(40)\\
7.45& 3.752 & 5450 & 12.4169 &1.927 & 0.079 & 31.69 &0.464 &0.07 &0.42 &500 &600 &5.5981(40)\\
7.45& 3.752 & 5400 & 12.8257 &1.936 & 0.105 & 44.08 &0.634 &0.09 &0.54 &700 &850 &8.6258(40)\\
7.45& 3.752 & 5300 & 13.7265 &1.955 & 0.148 & 61.86 &0.855 &0.13 &0.76 &1000 &1200 &2.3433(41)\\
7.45& 3.752 & 5200 & 14.6879 &1.972 & 0.174 & 69.28 &0.951 &0.15 &0.93 &1100 &1400 &3.6237(41)\\
7.45& 3.752 & 5100 & 15.6774 &1.990 & 0.196 & 71.38 &1.011 &0.17 &1.01 &1250 &1500 &5.4076(41)\\
7.45& 3.752 & 5000 & 16.7485 &2.005 & 0.196 & 67.49 &0.964 &0.17 &0.91 &1150 &1400 &7.9229(41)\\
7.45& 3.752 & 4900 & 17.7920 &2.015 & 0.136 & 43.37 &0.520 &0.12 &0.52 &650 &800 &2.7150(41)\\
\hline\noalign{\smallskip}\noalign{\smallskip} 
\end{tabular}
\begin{minipage}[h]{18.05cm} 
$^a$ Stellar mass (solar units).  
\hspace*{0.5mm} $^b$ Logarithmic luminosity (solar units).  
\hspace*{0.5mm} $^c$ Effective temperature (K).
\hspace*{0.5mm} $^d$ Fundamental period (day). 
\hspace*{0.5mm} $^e$ Logarithmic mean radius (solar units). 
\hspace*{0.5mm} $^f$ Fractional radius variation.  
\hspace*{0.5mm} $^g$ Radial velocity amplitude (Km$s^{-1}$).
\hspace*{0.5mm} $^h$ Bolometric amplitude (mag).
\hspace*{0.5mm} $^i$ Amplitude of logarithmic static gravity.  
\hspace*{0.5mm} $^j$ Amplitude of logarithmic effective gravity.
\hspace*{0.5mm} $^k$ Surface temperature variation (K).   
\hspace*{0.5mm} $^l$ Effective temperature variation (K).
\hspace*{0.5mm} $^m$ Total kinetic energy (erg).
\end{minipage} 
\end{table}
\end{flushleft}


\begin{flushleft} 
\begin{table}
\begin{tabular}{cccccccc}
\multicolumn{8}{c}{Table 2: HP Systematic Behavior}\\  
\noalign{\smallskip}\noalign{\smallskip}\hline\noalign{\smallskip}  
$M^a$&
$T_{\rm e}^b$&
$P_2/P_0^c$& 
$A(V)^d$& 
$Ac(M_{\rm V})^e$& 
$Sk(M_{\rm V})^f$&
$Ac(V_{\rm R})^g$&  
$Sk(V_{\rm R})^h$\\
(1)&
(2)&
(3)&
(4)&
(5)&
(6)&
(7)&
(8)\\  
\hline\noalign{\smallskip}
6.55& 5900 &0.539 &0.735 &2.378$^i$ &1.387$^i$ &2.058$^i$ &0.923$^i$\\ 
6.55& 5800 &0.537 &0.944 &2.717 &1.123 &2.953 &0.852 \\  
6.55& 5700 &0.533 &0.905 &2.559 &0.721 &2.717 &0.789 \\  
6.55& 5600 &0.528 &0.794 &2.021 &0.475 &2.378 &0.600 \\  
6.55& 5500 &0.526 &0.682 &1.755 &0.453 &1.976 &0.572  \\  
6.55& 5400 &0.525 &0.551 &1.976 &0.460 &2.021 &0.605 \\  
6.55& 5300 &0.522 &0.350 &2.759 &0.481 &2.021 &0.592 \\   
6.55& 5250 &0.520 &0.244 &2.774 &0.468 &1.915 &0.550\\   
6.55& 5200 &0.517 &0.187 &0.546 &0.471 &0.650 &0.543  \\   
6.55& 5150 &0.514 &0.210 &0.587 &0.548 &0.681 &0.706 \\  
6.55& 5100 &0.514 &0.265 &0.647 &0.618 &0.730 &0.862 \\  
6.70& 5900 &0.535 &0.476 &1.924 &1.222 &1.710 &0.945\\  
6.70& 5800 &0.532 &0.879 &2.774 &1.150 &2.663 &0.828\\  
6.70& 5700 &0.530 &0.886 &2.390 &0.689 &2.584 &0.739 \\  
6.70& 5600 &0.525 &0.797 &1.959 &0.508 &2.268 &0.577\\  
6.70& 5500 &0.523 &0.687 &1.915 &0.490 &2.077 &0.582\\  
6.70& 5400 &0.522 &0.514 &2.165 &0.479 &2.155 &0.600\\  
6.70& 5350 &0.524 &0.393 &2.802 &0.473 &2.096 &0.582\\  
6.70& 5300 &0.519 &0.299 &2.846 &0.460 &2.067 &0.538 \\  
6.70& 5250 &0.518 &0.248 &0.538 &0.473 &0.629 &0.515 \\  
6.70& 5200 &0.515 &0.299 &0.608 &0.527 &0.658 &0.615\\
6.70& 5100 &0.510 &0.485 &0.818 &0.681 &0.818 &0.965 \\  
6.85& 5800 &0.530 &0.819 &2.484 &1.141 &2.677 &0.815 \\  
6.85& 5700 &0.525 &0.864 &2.279 &0.704 &2.496 &0.647 \\  
6.85& 5600 &0.523 &0.795 &2.021 &0.550 &2.247 &0.580 \\  
6.85& 5500 &0.522 &0.690 &2.003 &0.517 &2.236 &0.585 \\  
6.85& 5450 &0.520 &0.583 &2.067 &0.501 &2.247 &0.590 \\  
6.85& 5400 &0.521 &0.473 &2.289 &0.479 &2.257 &0.575 \\  
6.85& 5350 &0.518 &0.362 &0.585 &0.466 &2.185 &0.531 \\  
6.85& 5300 &0.516 &0.311 &0.580 &0.481 &0.626 &0.497 \\  
6.85& 5250 &0.513 &0.367 &0.647 &0.527 &0.664 &0.577 \\  
6.85& 5200 &0.511 &0.547 &0.773 &0.577 &0.761 &0.718\\  
6.85& 5100 &0.506 &0.804 &1.075 &0.727 &0.972 &1.024 \\  
7.00& 5800 &0.528 &0.713 &2.110 &1.078 &2.365 &0.801 \\
7.00& 5700 &0.525 &0.829 &2.154 &0.733 &2.437 &0.621 \\
7.00& 5600 &0.522 &0.777 &1.985 &0.587 &2.279 &0.580\\
7.00& 5500 &0.524 &0.622 &2.096 &0.536 &2.311 &0.587 \\
7.00& 5450 &0.515 &0.517 &2.030 &0.499 &2.333 &0.567 \\
7.00& 5400 &0.516 &0.388 &0.597 &0.486 &2.205 &0.524 \\
7.00& 5350 &0.513 &0.375 &0.613 &0.501 &0.631 &0.488 \\
7.00& 5300 &0.513 &0.446 &0.675 &0.536 &0.681 &0.577 \\
7.00& 5200 &0.507 &0.858 &1.042 &0.885 &0.939 &0.866 \\
7.00& 5100 &0.502 &1.065 &1.317 &0.959 &1.113 &1.072 \\
7.00& 5000 &0.500 &0.868 &1.278 &0.977 &1.189 &1.257 \\
\hline\noalign{\smallskip}
\end{tabular}
\end{table}
\end{flushleft}

\begin{flushleft} 
\begin{table}
\begin{tabular}{cccccccc}
\multicolumn{8}{c}{Table 2: Continued}\\  
\noalign{\smallskip}\noalign{\smallskip}\hline\noalign{\smallskip}  
$M^a$&
$T_{\rm e}^b$&
$P_2/P_0^c$& 
$A(V)^d$& 
$Ac(M_{\rm V})^e$& 
$Sk(M_{\rm V})^f$&
$Ac(V_{\rm R})^g$&  
$Sk(V_{\rm R})^h$\\
(1)&
(2)&
(3)&
(4)&
(5)&
(6)&
(7)&
(8)\\  
\hline\noalign{\smallskip}
7.15& 5800 &0.525 &0.568 &1.809 &1.024 &1.994 &0.792\\  
7.15& 5700 &0.522 &0.738 &2.215 &0.754 &2.195 &0.626 \\  
7.15& 5600 &0.518 &0.720 &2.086 &0.631 &2.378 &0.590 \\  
7.15& 5550 &0.517 &0.631 &2.077 &0.585 &2.356 &0.582 \\  
7.15& 5500 &0.517 &0.546 &1.994 &0.534 &2.401 &0.560 \\  
7.15& 5450 &0.514 &0.445 &0.645 &0.508 &2.333 &0.515\\  
7.15& 5400 &0.513 &0.434 &0.658 &0.524 &0.647 &0.497\\  
7.15& 5350 &0.513 &0.543 &0.742 &0.565 &0.712 &0.587 \\  
7.15& 5300 &0.508 &0.802 &0.919 &0.972 &0.855 &0.692 \\  
7.15& 5200 &0.502 &1.130 &1.299 &0.965 &1.114 &0.927 \\  
7.15& 5100 &0.498 &1.274 &1.538 &1.141 &1.247 &1.392 \\  
7.15& 5000 &0.497 &1.159 &1.525 &1.457 &1.469 &1.358 \\  
7.15& 4900 &0.498 &0.507 &1.252 &1.092 &1.174 &1.146 \\  
7.30& 5800 &0.524 &0.449 &1.577 &0.996 &1.747 &0.802 \\  
7.30& 5700 &0.522 &0.716 &2.049 &0.795 &2.311 &0.621\\ 
7.30& 5600 &0.516 &0.647 &2.067 &0.647 &2.356 &0.577 \\  
7.30& 5575 &0.519 &0.635 &2.154 &0.610 &2.413 &0.575\\  
7.30& 5550 &0.516 &0.585 &2.067 &0.587 &2.413 &0.558\\  
7.30& 5525 &0.514 &0.532 &2.030 &0.555 &2.425 &0.541 \\  
7.30& 5500 &0.516 &0.493 &0.701 &0.536 &2.413 &0.524 \\  
7.30& 5475 &0.514 &0.475 &0.683 &0.538 &2.367 &0.501 \\  
7.30& 5450 &0.513 &0.471 &0.678 &0.543 &0.653 &0.495 \\  
7.30& 5400 &0.511 &0.562 &0.739 &0.570 &0.704 &0.555 \\  
7.30& 5300 &0.510 &1.013 &1.101 &1.028 &0.992 &0.779\\  
7.30& 5200 &0.500 &1.271 &1.433 &0.980 &1.237 &0.953 \\  
7.30& 5100 &0.495 &1.351 &1.611 &1.299 &1.404 &1.427 \\  
7.30& 5000 &0.494 &1.291 &1.793 &1.778 &1.625 &1.375 \\  
7.30& 4900 &0.478 &0.345 &1.146 &1.079 &1.110 &1.092\\  
7.45& 5800 &0.522 &0.178 &1.283 &0.972 &1.273 &0.894\\  
7.45& 5700 &0.530 &0.628 &1.994 &0.789 &2.215 &0.618\\  
7.45& 5625 &0.516 &0.585 &2.106 &0.664 &2.344 &0.572 \\  
7.45& 5600 &0.515 &0.556 &2.125 &0.623 &2.344 &0.558\\  
7.45& 5575 &0.514 &0.513 &2.049 &0.582 &2.356 &0.543\\  
7.45& 5550 &0.513 &0.487 &0.727 &0.560 &2.356 &0.522 \\
7.45& 5525 &0.512 &0.487 &0.724 &0.567 &2.356 &0.499 \\  
7.45& 5500 &0.511 &0.500 &0.733 &0.580 &0.667 &0.504 \\  
7.45& 5450 &0.508 &0.616 &0.782 &0.615 &0.724 &0.565 \\  
7.45& 5400 &0.507 &0.854 &0.923 &0.631 &0.852 &0.653 \\  
7.45& 5300 &0.501 &1.195 &1.252 &0.992 &1.155 &0.848 \\  
7.45& 5200 &0.497 &1.372 &1.538 &1.028 &1.347 &1.304 \\  
7.45& 5100 &0.494 &1.492 &1.774 &1.389 &1.699 &1.463 \\  
7.45& 5000 &0.489 &1.419 &2.378 &1.740 &1.770 &1.381 \\  
7.45& 4900 &0.489 &0.823 &1.710 &1.212 &1.433 &1.207 \\
\hline\noalign{\smallskip}\noalign{\smallskip} 
\end{tabular}
\begin{minipage}[h]{11.75cm} 
$^a$ Stellar mass (solar units).  
\hspace*{0.5mm} $^b$ Effective temperature (K).
\hspace*{0.5mm} $^c$ Second overtone to fundamental period ratio. 
\hspace*{0.5mm} $^d$ Visual amplitude (mag). 
\hspace*{0.5mm} $^e$ Acuteness of visual light curve.  
\hspace*{0.5mm} $^f$ Skewness of visual light curve.
\hspace*{0.5mm} $^g$ Acuteness of radial velocity curve.
\hspace*{0.5mm} $^h$ Skewness of radial velocity curve.
\hspace*{0.5mm} $^i$ The $Ac$ and $Sk$ values based on light curves were 
estimated by adopting the magnitude scale -the maximum is the minimum value-, 
while the $Ac$ and $Sk$ values based on velocity curves were estimated by 
adopting the theoretical notation i.e. the maximum value is attained during 
the outward excursion (positive values). Both for light and velocity curves
we assumed that the phase is zero at maximum.
\end{minipage}
\end{table}
\end{flushleft}

\begin{flushleft} 
\begin{table}
\begin{tabular}{lcc}
\multicolumn{3}{c}{Table 3: Input parameters and nonlinear results.}\\  
\noalign{\smallskip}\noalign{\smallskip}\hline\noalign{\smallskip}  
Parameter$^a$&
SK$^b$&
III$^c$ \\  
(1)&
(2)&
(3)\\ 
\hline\noalign{\smallskip}\noalign{\smallskip}  
 $M$             & 5.35   & 7.00 \\
 $log L$         & 3.65   & 3.65 \\  
  $T_{\rm e}$          & 5700   & 5700 \\  
   Y             & 0.29   & 0.25 \\  
   Z             & 0.01   & 0.008 \\  
  $P_0$          &11.0398 & 9.1025 \\   
$P_2/P_0$        &0.4937  & 0.5249 \\   
$\Delta u$       & 37.8   & 38.8 \\ 
$\Delta M_{{\rm {bol}}}$ & 0.71   & 0.65 \\ 
\hline\noalign{\smallskip}\noalign{\smallskip}  
\end{tabular}
\begin{minipage}[h]{7.2cm} 
$^a$ Y and Z are the helium and the metal content respectively.
The other symbols and their units are the same as in Tables 1 and 2. 
\hspace*{0.5mm} $^b$ SK model. 
\hspace*{0.5mm} $^c$ Paper III model.  
\end{minipage}
\end{table}
\end{flushleft}